\documentclass{aastex631}
\usepackage{amsmath}
\usepackage{gensymb}
\usepackage{float}
\usepackage{wrapfig}
\usepackage{makecell}

\received{August 19, 2022}
\revised{April 22, 2023}
\accepted{April 24, 2023}
\submitjournal{Icarus}

\shorttitle{Drift Rates of Major Neptunian Features}
\shortauthors{Chavez et al.}

\graphicspath{{./}{figures/}}

\begin{document}

\title{Drift Rates of Major Neptunian Features between 2018 and 2021}

\author{Erandi Chavez}
\affiliation{Center for Astrophysics $|$ Harvard \& Smithsonian, 60 Garden St, Cambridge, MA 02138, USA}

\author{Erin Redwing}
\affiliation{Department of Earth and Planetary Science, 307 McCone Hall, University of California, Berkeley, CA 94720, USA}

\author{Imke de Pater}
\affiliation{Department of Astronomy, 501 Campbell Hall, 
University of California, Berkeley, CA 94720, USA}
\affiliation{Department of Earth and Planetary Science, 307 McCone Hall, University of California, Berkeley, CA 94720, USA}

\author{Ricardo Hueso}
\affiliation{Departamento Física Aplicada I, Escuela Ingeniería de Bilbao, Universidad del País Vasco
UPV/EHU, Spain}

\author{Edward M. Molter}
\affiliation{Department of Astronomy, 501 Campbell Hall, 
University of California, Berkeley, CA 94720, USA}

\author{Michael H. Wong}
\affiliation{Department of Astronomy, 501 Campbell Hall, 
University of California, Berkeley, CA 94720, USA}

\author{Carlos Alvarez}
\affiliation{W. M. Keck Observatory, 65-1120 Mamalahoa Hwy, Kamuela HI 96743, USA}

\author{Elinor Gates}
\affiliation{Lick Observatory, P.O. Box 85, Mount Hamilton, CA 95140, USA}

\author{Katherine de Kleer}
\affiliation{Division of Geological and Planetary Sciences, California Institute of Technology, Pasadena, CA 91125, USA}

\author{Joel Aycock}
\affiliation{W. M. Keck Observatory, 65-1120 Mamalahoa Hwy, Kamuela HI 96743, USA}

\author{Jason Mcilroy}
\affiliation{The Stratospheric Observatory for Infrared Astronomy (SOFIA), NASA Armstrong Flight Research Center (AFRC), Palmdale, CA}

\author{John Pelletier}
\affiliation{W. M. Keck Observatory, 65-1120 Mamalahoa Hwy, Kamuela HI 96743, USA}

\author{Anthony Ridenour}
\affiliation{W. M. Keck Observatory, 65-1120 Mamalahoa Hwy, Kamuela HI 96743, USA}

\author{Agustín Sánchez-Lavega}
\affiliation{Departamento Física Aplicada I, Escuela Ingeniería de Bilbao, Universidad del País Vasco
UPV/EHU, Spain}

\author{Jose Félix Rojas}
\affiliation{Departamento Física Aplicada I, Escuela Ingeniería de Bilbao, Universidad del País Vasco
UPV/EHU, Spain}

\author{Terry Stickel}
\affiliation{W. M. Keck Observatory, 65-1120 Mamalahoa Hwy, Kamuela HI 96743, USA}

\begin{abstract}

Using near-infrared observations of Neptune from the Keck and Lick Observatories, and the Hubble Space Telescope in combination with amateur datasets, we calculated the drift rates of prominent infrared-bright cloud features on Neptune between 2018 and 2021. These features had lifespans of $\sim 1$ day to $\geq$1 month and were located at mid-latitudes and near the south pole. Our observations permitted determination of drift rates via feature tracking. These drift rates were compared to three zonal wind profiles describing Neptune’s atmosphere determined from features tracked in H band (1.6 $\mu m$), K’ band (2.1 $\mu m$), and Voyager 2 data at visible wavelengths. Features near $-70 \degree$ measured in the F845M filter (845nm) were particularly consistent with the K' wind profile. The southern mid-latitudes hosted multiple features whose lifespans were $\geq$1 month, providing evidence that these latitudes are a region of high stability in Neptune’s atmosphere. We also used HST F467M (467nm) data to analyze a dark, circumpolar wave at $- 60 \degree$ latitude observed on Neptune since the Voyager 2 era. Its drift rate in recent years (2019-2021) is $4.866 \pm 0.009 \degree $/day. This is consistent with previous measurements by \cite{karkoschka2011}, which predict a $4.858 \pm 0.022 \degree$/day drift rate during these years. It also gained a complementary bright band just to the north.

\end{abstract}


\section{Introduction} \label{sec:intro}

Neptune’s small angular size (2.3”) has made high resolution images of the planet historically difficult to achieve. The first detailed images of Neptune were taken during Voyager 2’s 1989 flyby of Neptune, which lasted 6 months \citep{smith1989}. These images led to many fundamental discoveries about the planet, including the first measurements of the planet's zonal wind speeds \citep{smith1989,limayeandsromovsky1991}, which revealed especially active and turbulent winds and sparked interest in the processes that control Neptune’s atmosphere. Following the Voyager 2 era, high-resolution observations of Neptune at visible and infrared wavelengths continued with the Hubble Space Telescope (HST) and ground-based telescopes that utilize adaptive optics (AO) such as the Canada-France-Hawaii Telescope (CFHT) and the W.M. Keck II Telescope \citep{roddier1997,max2003,gibbard2003,hammelandlockwood1997,sromovsky2001c}. AO technology corrected for the distortion caused by Earth's atmosphere, which allowed high-resolution observations to continue with ground-based telescopes. 

From the Voyager 2 data, \cite{limayeandsromovsky1991} tracked the zonal velocities of various cloud features and \cite{sromovsky1993} calculated a zonal wind profile based on a polynomial fit of these data after binning them using $1 \degree$ latitude bins. All data were obtained at visible wavelengths, where Neptune’s disk appears opaque and hazy. \cite{karkoschka2011} later calculated additional wind speeds from these same data, and the Voyager 2 zonal wind profile was updated to a Fourier cosine series fit by \cite{sanchez-lavega2019}, which we use as the canonical Voyager 2 wind profile. Recent zonal wind profiles have been created using near-infrared (near-IR) data with the Near InfraRed Camera 2’s (NIRC2) H filter \citep{martin2012,fitzpatrick2014,tollefson2018} and K’ filter \citep{fitzpatrick2014,tollefson2018} using the W.M. Keck II telescope with AO. Unlike at visible wavelengths, in these near-IR filters Neptune is characterized by a dark background contrasted by bright cloud features, typically located at mid-latitudes. This contrast, combined with the high quality of near-IR images produced by using AO, allows for detailed observation of Neptune’s atmosphere and clouds. 

\begin{figure}
    \centering
    \includegraphics[width=0.95\textwidth]{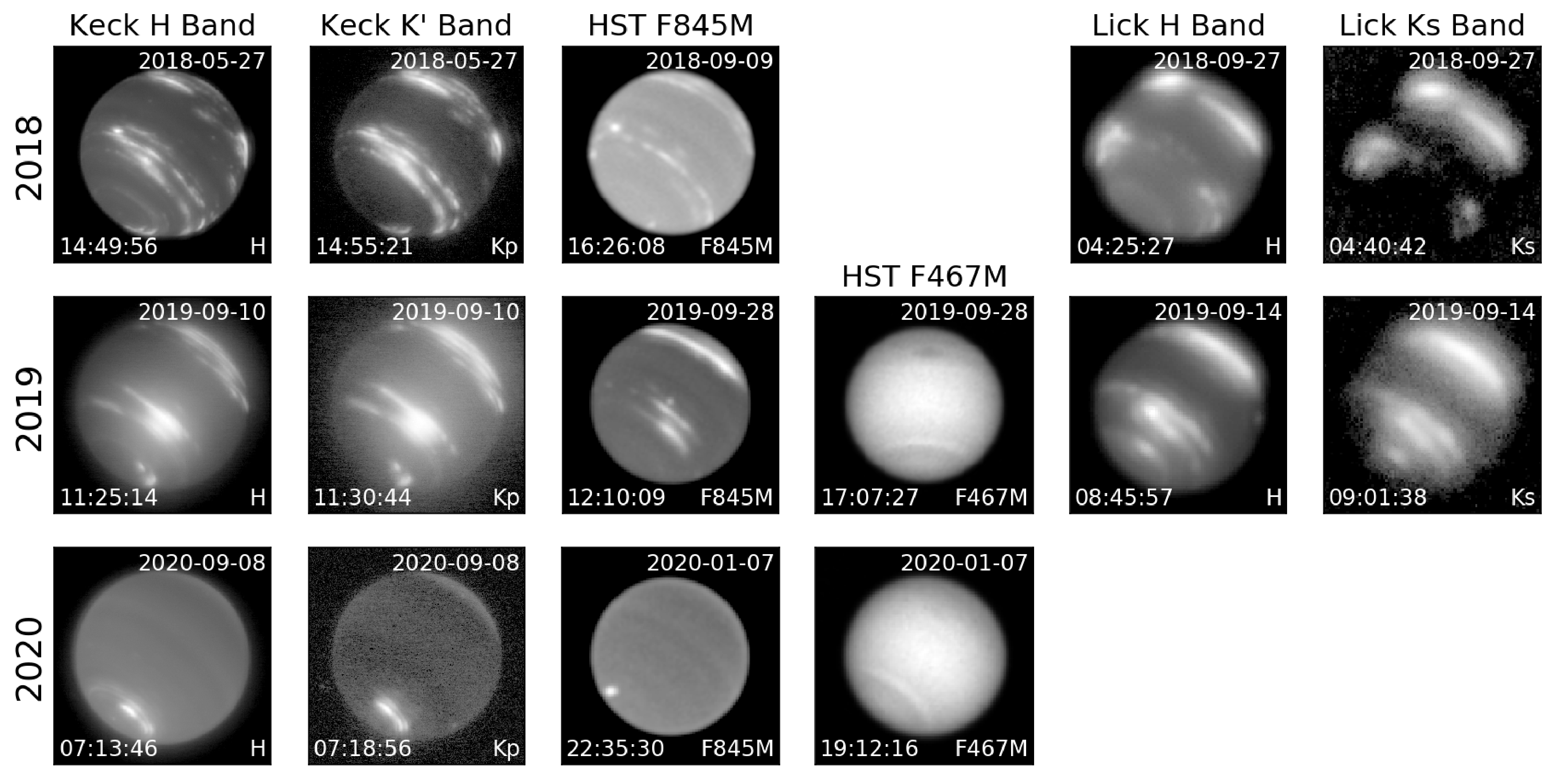}
    \caption{From left to right, representative Keck H band, Keck K’ band, HST F845M, HST F467M, Lick H band, and Lick Ks band images during 2018 (top row) 2019 (middle row) and 2020 (bottom row). F467M images are displayed linearly, while all other images are shown logarithmically}. Neptune's low cloud activity during 2020 meant we increased the image stretch to see the disk clearly in H and K' band. Similarly, we also increased the image stretch for the Lick Ks band data, which led to grainy images. 
    \label{fig:example_image_grid}
\end{figure}

In this paper, we present observations of Neptune’s atmosphere from 2018 to 2021 from various telescopes, including the Keck Observatory in H and K’ band, the Shane Telescope at the Lick Observatory in H and Ks band, the Hubble Space Telescope in FQ619N (619 nm), F657N (657 nm), FQ727N (727 nm), F763M (763 nm) and F845M (845nm), amateur astronomer data, and others. We perform cloud tracking to determine the drift rates of major cloud features identified between 2018 and 2021. While most features are tracked in a single filter (H band, K' band, Ks band, and F845M), the longest-lived cloud features ($\geq$ 1 month) are tracked using multiple near-IR filters from many different telescopes---including amateur astronomer data---to improve time coverage of the feature. We compare the drift rates of these various features to the canonical Voyager 2 wind profile \citep{sromovsky1993} updated by \cite{sanchez-lavega2019} with results from \cite{karkoschka2011}, and the H and K’ band wind profiles from \cite{tollefson2018} that were determined by tracking cloud features observed in H and K' band data, respectively. 

Using data from HST's F467M filter (467 nm) at blue wavelengths, we also analyze a dark wave near Neptune's South Pole, the South Polar Wave (SPW). It is notable as Neptune's most stable feature, having persisted since its initial observation in Voyager 2 data \citep{smith1989,sromovsky2001a,sromovsky2001b,karkoschka2011}. The dark wave is contrasted against Neptune's brighter surrounding background in F467M. It also has connections with the South Polar Feature (SPF), a group of small, bright, and rapidly-evolving cloud features between $-68 \degree$ and $-75 \degree$ latitude that were originally observed in Voyager 2 data \citep{limayeandsromovsky1991,sromovsky1993}. When the short-lived individual cloud features are tracked over several hours, the SPF periods range between 11 and 13 hrs \citep{sromovsky1993,martin2012}, however when tracked as a single group or over many years, the SPF has an apparent rotation rate of 15.9663 hr (initially measured by \citealt{hammel1989}, updated by \citealt{sromovsky1993}, and then by \citealt{karkoschka2011}). The South Polar Feature and the South Polar Wave had identical long-term periods of 15.9663 hr when observed between 1989 and 2010, resulting in a rotational lock between the two that lasted at least two decades \citep{karkoschka2011}. We analyze the SPW's drift rate in recent years, comparing it to the previous result of $5 \degree$/day determined by \cite{karkoschka2011} and highlighting possible connections to recent near-IR cloud features observed at the south pole.

We describe the data acquisition and reduction procedures in Section \ref{sec:data}, the tracking process of prominent features in Section \ref{sec:feature_tracking}, and the tracking process for the Southern Polar Wave in Section \ref{sec:bright_ring_dark_band}. We discuss the results and our conclusions in Sections \ref{sec:discussion} and \ref{sec:conclusions}.

\section{Data Description} \label{sec:data}
\begin{deluxetable}{ccccc}
\tablecaption{List and Description of Telescope Filters \label{tab:telescope_filters}}
\tablewidth{0pt}
\tablehead{
\colhead{Telescope} & \colhead{Band} & \colhead{Central Wavelength $\lambda$ ($\mu m$)} & \colhead{Range $\Delta \lambda$ ($\mu m$)} & \colhead{Number of Observations}
}
\startdata
Keck NIRC2 & H & 1.63 & 0.30 & 11 \\
Keck NIRC2 & K' & 2.12 & 0.35 & 9 \\
Lick ShARCS & H & 1.66 & 0.30 & 1 \\
Lick ShARCS & Ks & 2.15 & 0.32 & 1 \\
HST WFPC2 & F467M & 0.4663 & 0.01664 & 8 \\
HST WFC3 & F467M & 0.4675 & 0.0230 & 148 \\
HST WFC3 & FQ619N & 0.6194 & 0.0062 & 4 \\
HST WFC3 & F657N & 0.6573 & 0.0094 & 1\\
HST WFC3 & FQ727N & 0.7275 & 0.0064 & 2 \\
HST WFC3 & F763M & 0.763 & 0.0780 & 2 \\
HST WFC3 & F845M & 0.8454 & 0.0870 & 56 \\
Calar Alto PlanetCam & RG1000 & 1.35 & 0.6 & 16 \\
Gemini & Ks & 2.15 & 0.31 & 1 \\
GTC HiPERCAM & Z & 0.889 & 0.1163 & 8 \\
Amateur	& R + IR & 0.85 & & 1 \\
Amateur	& Red & 0.85 & & 1 \\
Amateur & R600 & 0.85 & & 3 \\
Amateur & IR785 & 0.85 & & 1 \\
\enddata
\end{deluxetable}

\subsection{Sources of Data}

\subsubsection{Keck and Lick}
Near-IR images were taken with the Near Infrared Camera 2 (NIRC2) on the Keck II Telescope in H band (1.63 $\mu m$) and K’ band (2.12 $\mu m$), and the Shane AO infraRed Camera and Spectrograph (ShARCS) with the Shane Telescope at Lick Observatory in H band (1.66 $\mu m$) and Ks band (2.15 $\mu m$) (see Table 1 for wavelength ranges of filters). The diffraction limit of the Keck II telescope at 2.1 $\mu m$ is $\sim 0.045$ arcseconds \citep{vandam2004} while the limit of the Shane Telescope at Lick observatory at 2.1 $\mu m$ is $\sim 0.15$ arcseconds. Both telescopes utilized adaptive optics (AO) systems. The Keck data spanned between 2018 and 2021, while the Lick data were taken between 2018 and 2020. These images were taken as part of the Twilight Zone\footnote{\url{https://www2.keck.hawaii.edu/inst/tda/TwilightZone}} program \citep{molter2019} which gives classically scheduled observers the opportunity to donate their unused telescope time (often during poor weather conditions or twilight hours) to take short (10 - 40 minute) observations of bright solar system objects. This program resulted in a high quantity of Keck and Lick images in which gaps between images ranged from days to months. The full list of images, including those from the Twilight program, can be found in Table \ref{tab:near_ir_tracking_images}. 

\subsubsection{HST}
The HST data were taken between 2018 and 2021 in multiple filters, however, we primarily used the F467M (467nm) and F845M (845 nm) images. Other filters (FQ619N, F657N, FQ727N F763M) were used as supplementary data in conjunction with the other telescope data described in Section \ref{sec:other_telescopes} to perform long-term feature tracking. The diffraction limit of HST at 0.6 $\mu m$ is $\sim 0.055$ arcseconds. HST images were taken on 2-4 days each year as part of the OPAL program \footnote{\url{https://archive.stsci.edu/prepds/opal/index.html}} \citep{simon2015} and as part of mid-cycles intended to observe the dark spot NGDS2018 \citep{simon2019,wong2022}. Each day of observation produced a dense sampling of images, with each frame taken minutes to hours apart during these days of observation. 

\subsubsection{Other Telescopes \label{sec:other_telescopes}}
\begin{figure}
    \centering
    \includegraphics[width=0.7\textwidth]{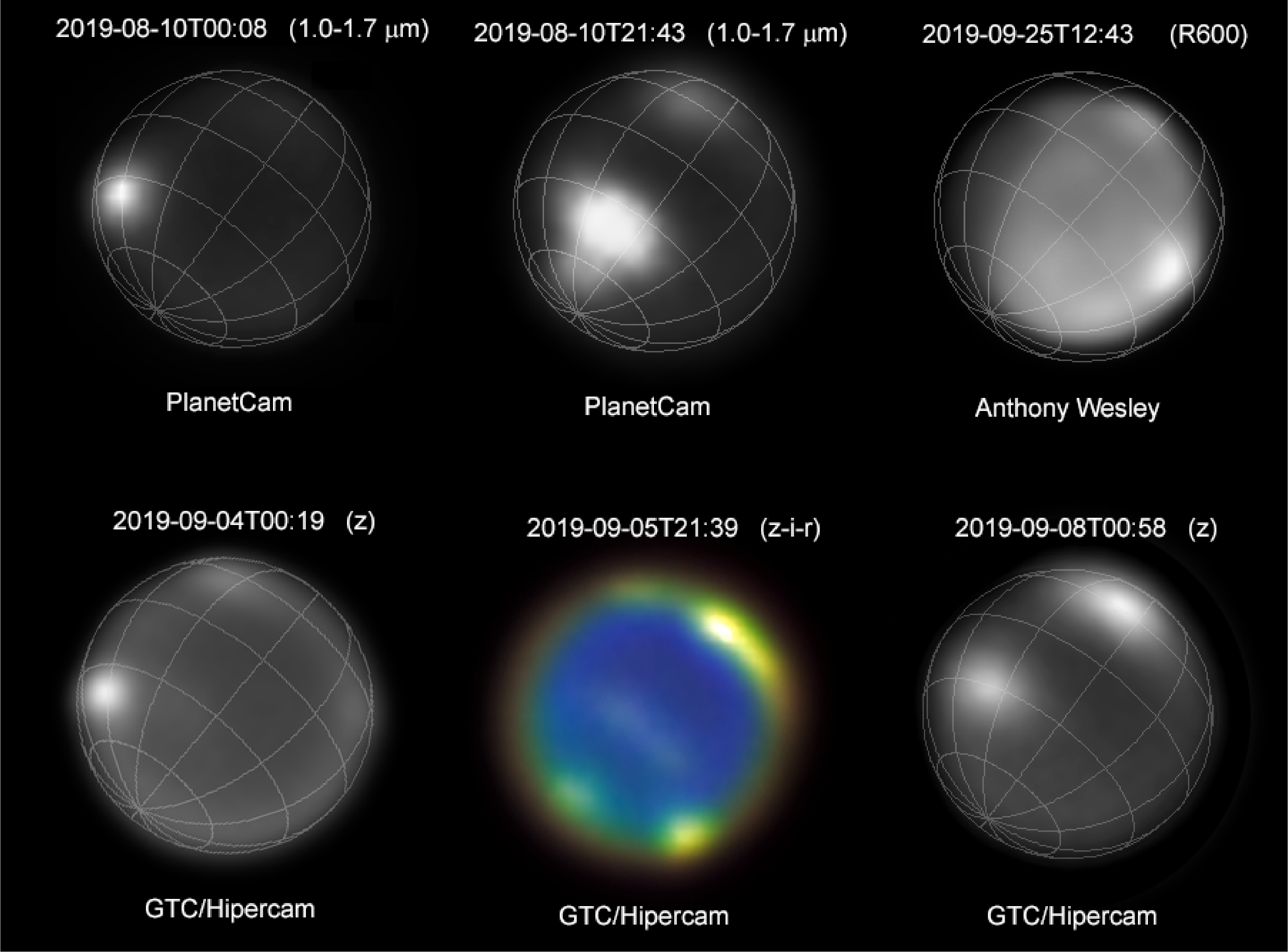}
    \caption{Examples of PlanetCam, amateur and GTC/HiPERCAM images over 2019. All images were navigated considering the size and position of Triton to identify correctly the size of the planet distorted from seeing. A thin grid shows longitudes and latitudes in intervals of 30 degrees except in the bottom-center image, which is a false color image composed from HiPERCAM images in the z-i-r filters. High-pass filters of different intensities were applied to every image.}
    \label{fig:example_pcam_grid}
\end{figure}

Images from various telescopes were taken during 2019 to supplement Keck and HST data for improved time coverage during long-term feature tracking. These include the PlanetCam instrument \citep{mendikoa2016} at the Calar Alto Observatory, the HiPERCAM instrument \citep{dhillon2021} at the Gran Telescopio Canarias (GTC), and the Gemini North Telescope. HiPERCAM and PlanetCam use the lucky imaging technique \citep{law2006} to improve the image quality over the atmospheric seeing and the Gemini North Telescope uses AO. The diffraction limits of these three telescopes at 2.1 $\mu m$ are $0.13 $ arcsec, $0.043$ arcsec, and $0.056$ arcsec, respectively. Different spatial resolutions over Neptune's disc were obtained on different nights. Three southern-latitude features during 2019 were particularly large and long-lived – they persisted for months on Neptune and were large and bright enough at red and near-IR wavelengths to be detected by the telescopes of amateur astronomers also using lucky imaging \cite{mousis2014}. With the addition of these large telescopes and amateur astronomer data, the time between observations of these 3 features was reduced to days instead of weeks. A list of filters associated with these telescopes is provided in Table 1; see Figure \ref{fig:example_pcam_grid} for an example of PlanetCam, GTC, and amateur astronomer data.

Amateur astronomers that obtained data of prominent features observed in Keck and HST during 2019 were R. Christensen, M. Delcroix, D. Millika, R. Sedrani, J. Sussenbach, and A. Wesley. They used telescopes with apertures of 12-14 inches and different filters in the red to infrared range, with the most frequent being a cut-off filter that starts at 785 nm. Most of these observations were retrieved from the PVOL database\footnote{\url{http://pvol2.ehu.eus/}} \citep{hueso2018}. Amateur data for later years were not frequent enough to secure positive identifications of real atmospheric features.

\subsection{Data Reduction}
For the data obtained by Keck and Lick telescopes, each frame underwent standard image reduction that involved sky subtraction, flat fielding, and median value masking to remove bad pixels. Data reduction was done in Python using the package nirc2\_reduce, which removed cosmic rays with the astroscrappy package \citep{mccully2018astroscrappy} and corrected the NIRC2 frames for the geometric distortion of the camera by applying the solution described in \cite{service2016}. A dither pattern with Neptune positioned in different regions of the detector was used to mitigate the effects of detector artifacts and to construct a sky background after median averaging the frames. A three-point dither was used for Keck observations and a five-point dither was used for Lick observations. After cropping and aligning the images, the frames were median averaged to produce a final image. Keck data were photometrically calibrated where possible using the methods described in \cite{chavez2022}.

HST images were taken by the UVIS detector of the Wide Field Camera 3 instrument and were reduced using standard HST reduction tools in the same manner as described in \cite{wong2018}. This included converting the data to units of I/F, as defined by \cite{hammel1989if}:
\begin{equation}
    \frac{I}{F} = \frac{r^2}{\Omega}\frac{F_{N}}{F_{\odot}}
\end{equation}

where $r$ is the ratio of Neptune and Earth's heliocentric distance in A.U., $\pi F_{\odot}$ is the Sun's Flux density at Earth's orbit \citep{colina1996}, $F_{N}$ is Neptune's observed flux density, and $\Omega$ is the solid angle corresponding to a single pixel on the detector. F467M images were corrected for limb-darkening using the Minnaert function \citep{minnaert1941}:

\begin{equation} \label{eq:1}
    I/F_{corr} = [I/F_{obs}] \mu_{0}^{-k}\mu^{1-k}
\end{equation}
$I/F_{obs}$ is the observed reflectivity, $\mu_0$ is the cosine of the solar incidence angle, and $\mu$ is the cosine of the emission angle. The constant k is empirically defined; we used k = 0.867 \citep{wong2018}.

\begin{figure}[t]
    \centering
    \includegraphics[width=1.0\textwidth]{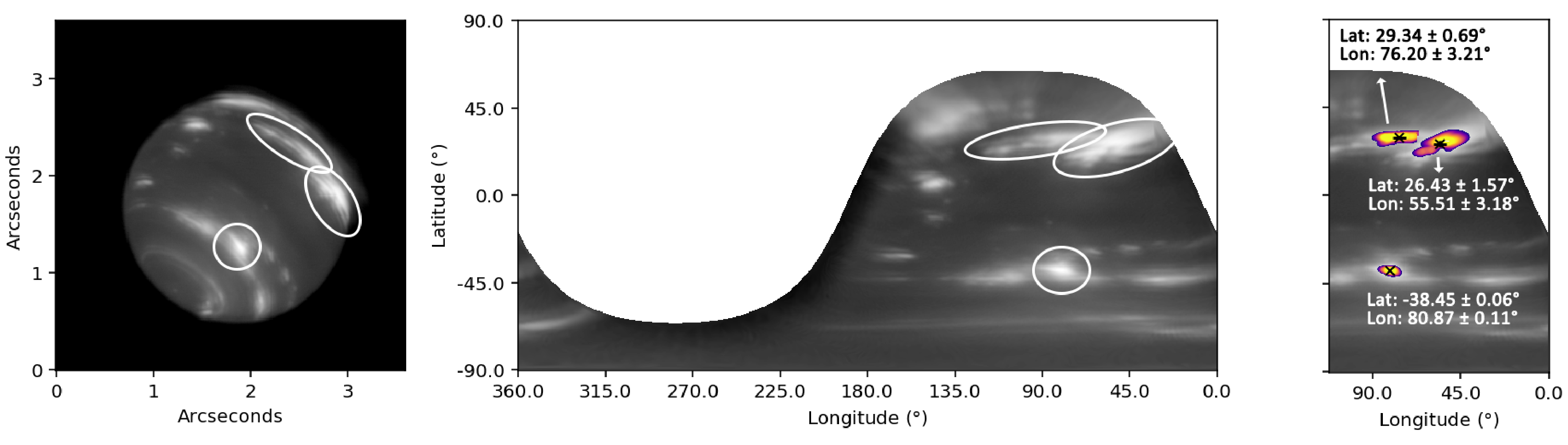}
    \caption{Process of selecting prominent features and calculating contours in a single image. Left: A Keck H band reduced image with prominent features circled in white. Center: The projected image corresponding to the left panel, with the same prominent features circled. Right: Contours for each prominent feature in the projected image are shown. The central positions of these prominent features are plotted and labeled.
}
    \label{fig:deproj_contours_ex}
\end{figure}

All Keck, Lick, and HST images were projected onto a flat latitude-longitude map (referred to as a ``projected image'') using the Python package “nirc2\_reduce” \cite[][see Figure \ref{fig:deproj_contours_ex} for an example of image projection]{nirc2reduce, molter2019} to facilitate assigning latitude/longitude coordinates to features on Neptune’s disk. All latitudes used in this paper are planetographic latitudes.

The PlanetCam, GTC, Gemini, and Amateur images were navigated in the program WinJupos (\url{http://jupos.org/gh/download.htm}), which uses Triton as an anchor point for Neptune’s latitude-longitude grid within an image. A detailed description of the program and reduction methods used for these data can be found in \cite{hueso2017}.

\section{Drift Rate Determination \label{sec:feature_tracking}}

\subsection{Locating Prominent Features}
\begin{figure}
    \centering
    \includegraphics[width=.99\textwidth]{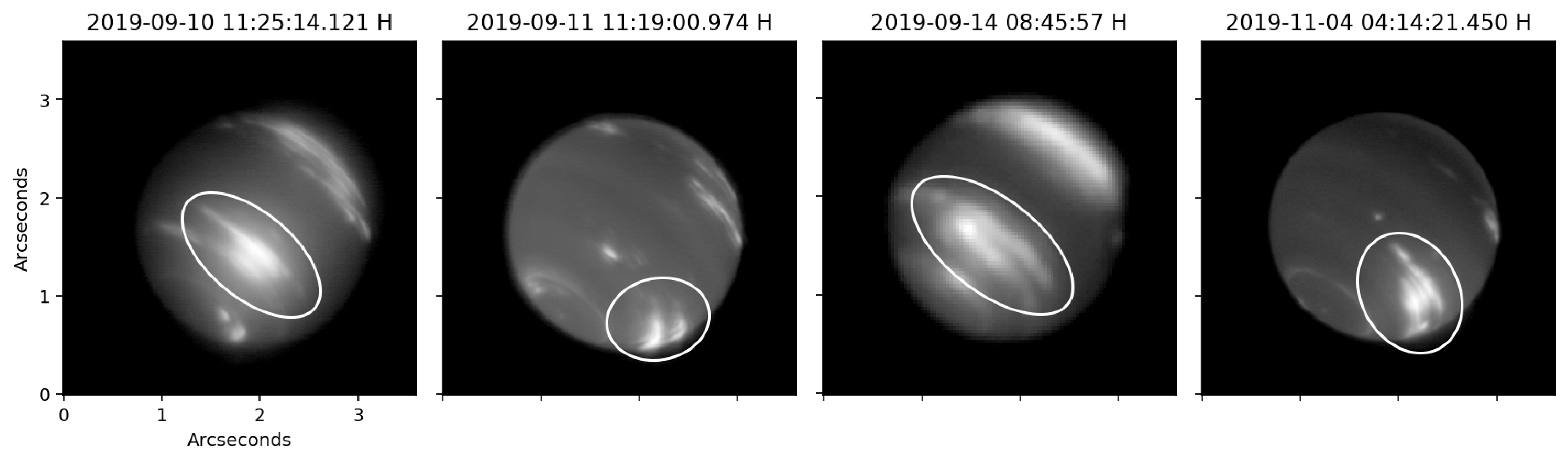}
    \caption{A demonstration of the feature tracking process over a single filter (H band). The same feature is shown in all four images, as determined by the criterion described in Section \ref{sec:same feature or not}.}
    \label{fig:tracking_example}
\end{figure}

Neptune's atmosphere is variable and small clouds break up in a short amount of time, typically in $<1$ hour time intervals \citep{limayeandsromovsky1991,sromovsky1993,martin2012}. The data we worked with, specifically the Lick and Keck data, were sporadic, with gaps between images varying between days and months apart. Taking this into account, we targeted large features that were easily identified between multiple images for cloud tracking. The brightest and/or largest clouds in an image were selected as prominent features. These prominent features typically spanned multiple degrees in latitude and longitude due to their size; however, the morphology between different prominent features was highly variable. Some features were compact with a clearly defined central location, while others were far larger, with complex morphologies that made the central point difficult to determine visually. This led us to use brightness contours to calculate the central point of the feature. 
We used the same method as in \cite{martin2012} and \cite{molter2019} to determine the central position of a prominent feature and its associated error: first, the feature was isolated within the projected image, making sure to contain the feature’s maximum while excluding any other features of similar or greater brightness. Next, multiple contours were generated within the isolated region, where each contour traced out a region in the image of equal brightness. The contours had values between 68\% and 95\% of the maximum value within the isolated region, with the contours occurring in intervals of .01\% for a total of 2700 contours. Using a different number of contours by increasing or reducing the interval size did not significantly affect the result, as shown in Figure \ref{fig:contours}. We calculated the centroid of each contour by finding the average latitude and longitude. The central position of the feature was determined to be the weighted average of these centroids, where each contour's centroid was weighted by the brightness of that contour:
\begin{equation}    
    x_c = \sum_{i=1}^{N} x_i I_i / {\sum_{i=1}^{N} I_i} \\
\end{equation}    
Here $x_i$ is the centroid of the $i^{\text{th}}$ contour, and ${x}_c$ is the weighted average of $N$ contours, each with a brightness $I_i$. ${x}_c$ is used as the central position of the feature. The standard deviation of the contour centroids was used as the 1-sigma error of the feature’s location. We repeated this process for every prominent feature within an image, and then for every image that contained prominent features.

\subsection{Same Feature or Not?\label{sec:same feature or not}}

Neptune’s dynamic atmosphere causes many features to shear apart, grow, or otherwise dramatically change in morphology over timescales of hours \citep{limayeandsromovsky1991}. To determine if a feature appeared in multiple images, individually tracked features had to be similar in latitude to each other, and the images needed to show the evolution of a feature’s morphology. A latitude band width of $\sim 5 \degree$ was typically used to correlate features between images, but the exact width of this latitude band depended on the feature in question. Tracking the evolution of the feature’s morphology was less quantitative. General characteristics between features needed to be consistent, such as size, brightness relative to the background, and similarity of the feature’s shape. If features in multiple images were in this latitude band and these images of the features showed the evolution of its morphology, then we assumed they were the same feature (see Figure \ref{fig:tracking_example} for an example).

To best compare a feature’s drift rate to the filter-specific H and K’ wind profiles mentioned in the Introduction (see also Figure \ref{fig:wind_speeds} discussed below), the above process was restricted to one filter at a time when possible. This resulted in most features being observed in these filters: H band, K' band, Ks band, or F845M. The HST F845M filter shows the largest contrast between clouds and the background atmosphere in HST images. This filter covers a strong methane absorption band, and it probes relatively similar altitudes in Neptune’s atmosphere, though not quite as high as the methane-sensitive K Band filters. Three specific long-lived features incorporated images from a multitude of telescopes and filters to increase time coverage. These three features were determined using the combination of HST data (all filters), Keck (H band), PlanetCam, GTC, Gemini, and amateur astronomer data.

\subsection{Cloud Tracking \label{sec:cloud_tracking}}

\begin{figure}[h]
    \centering
    \gridline{\fig{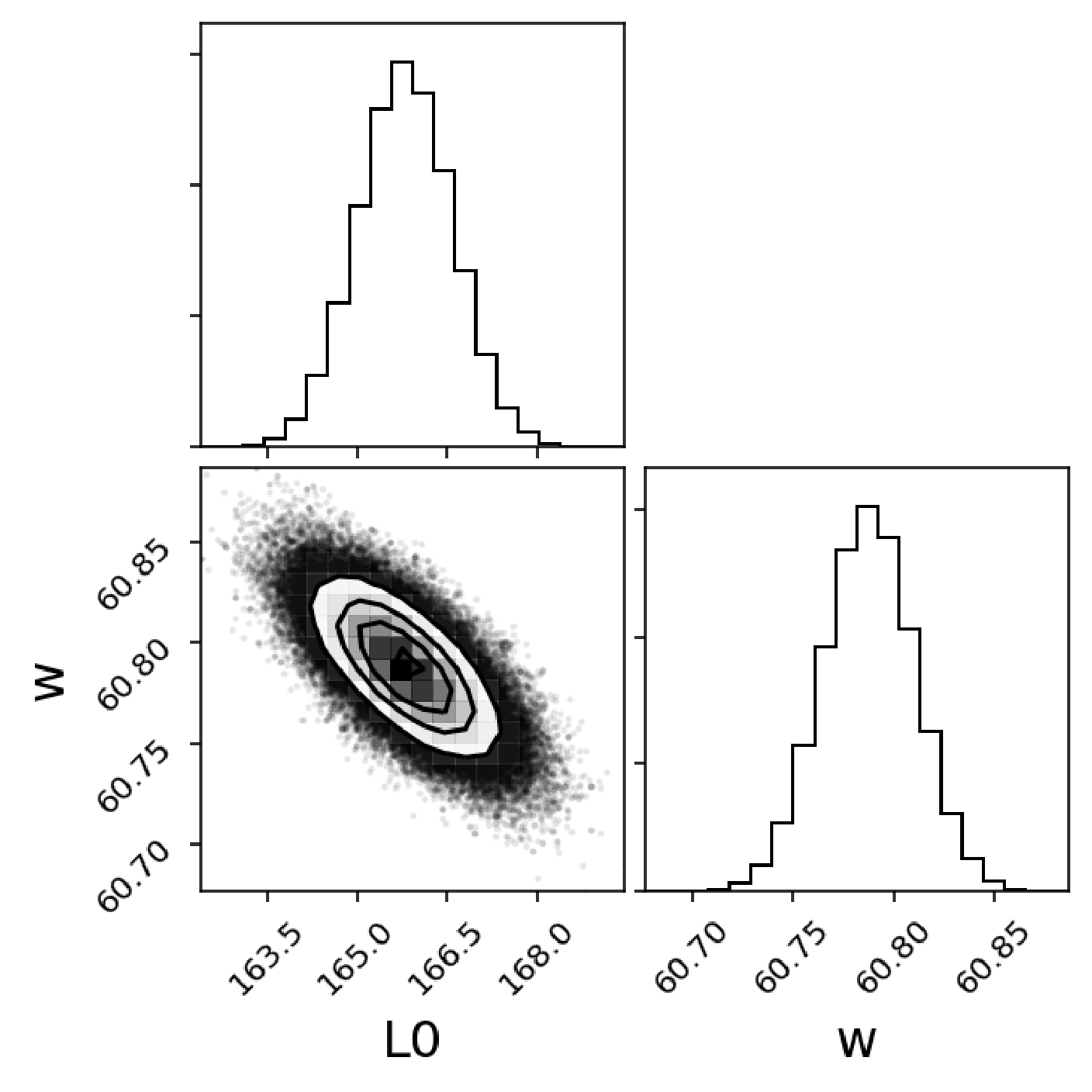}{0.49\textwidth}{}
          \fig{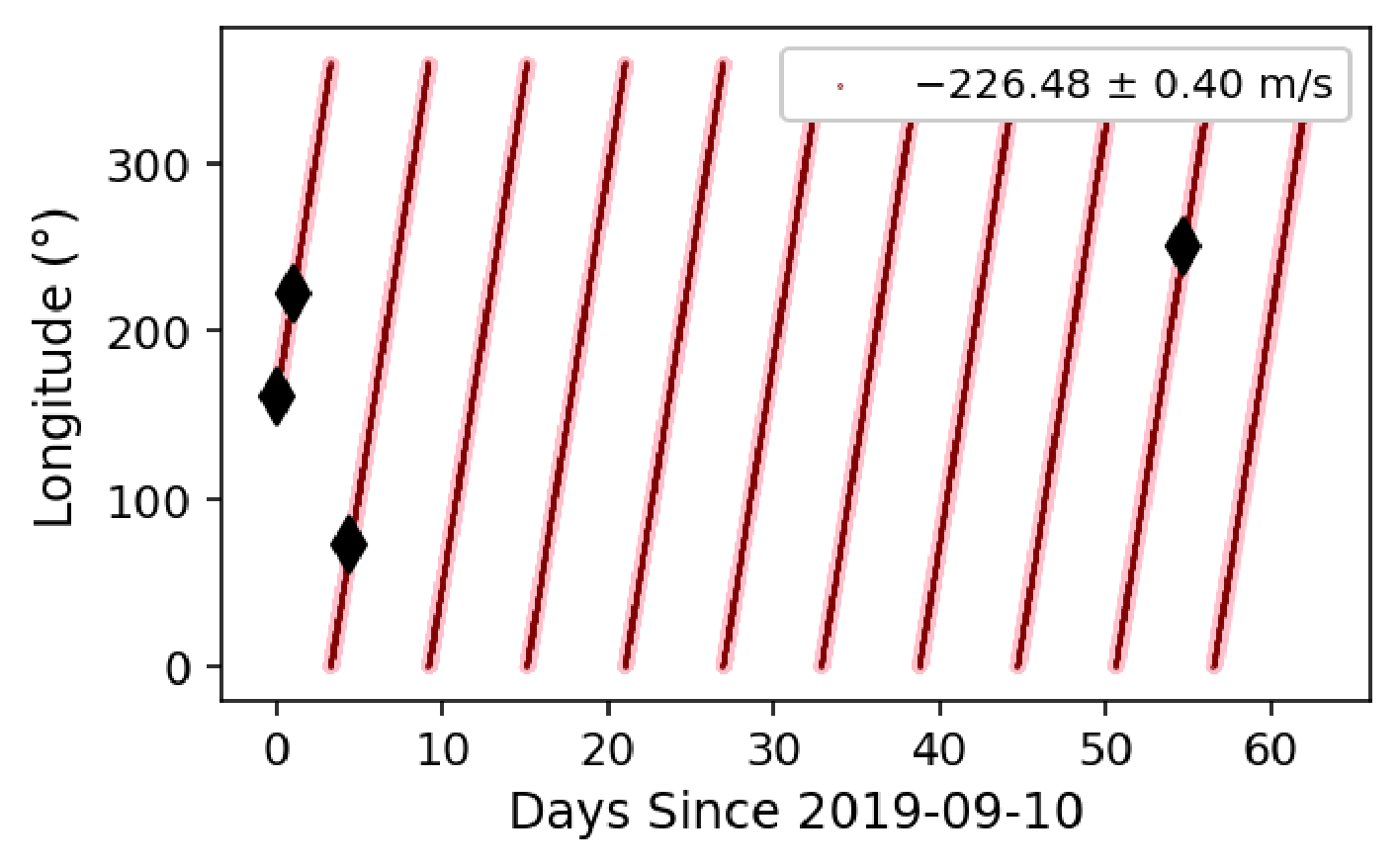}{0.49\textwidth}{}}
    \caption{Feature tracking results for the feature shown in Figure \ref{fig:tracking_example}. Left: Corner plots for the parameters $L0$ and $w$. The relevant parameter for our analysis is $w$, the drift rate (shown in degrees per day). Right: The data (black diamonds)} and feature tracking results are shown. The thin dark line represents the resulting eastward drift rate (50th percentile) in meters per second and the pink shaded region represents the drift rate error (84th and 16th percentiles).
    \label{fig:mcmc_chains_and_speed}
\end{figure}

After determining which images a particular feature appeared in, the next step was to determine the feature’s drift rate. In this paper, we are only concerned with longitudinal drift rate, and therefore all drift rates mentioned are zonal. In doing this, we assumed that every feature had a constant speed. To determine the speed, we fit a linear model to the feature’s longitudinal position as a function of time:
\begin{equation}
    L(t,t0,L0,w) = L0 + w(t-t0)
\end{equation}
$L0$ is the feature's initial longitude, $w$ is the drift rate, and $t0$ is the initial time of observation. All drift rates in this paper are listed in degrees per Earth day. The temporal gap between images meant that a single prominent feature could encircle Neptune many times during its observed lifetime. The model looped continuously between $0\degree$  and $360 \degree$ longitude in our calculations to account for this, as shown in Figure \ref{fig:mcmc_chains_and_speed}.

\begin{figure}[ht]
    \centering
    \includegraphics[width=1.0\textwidth]{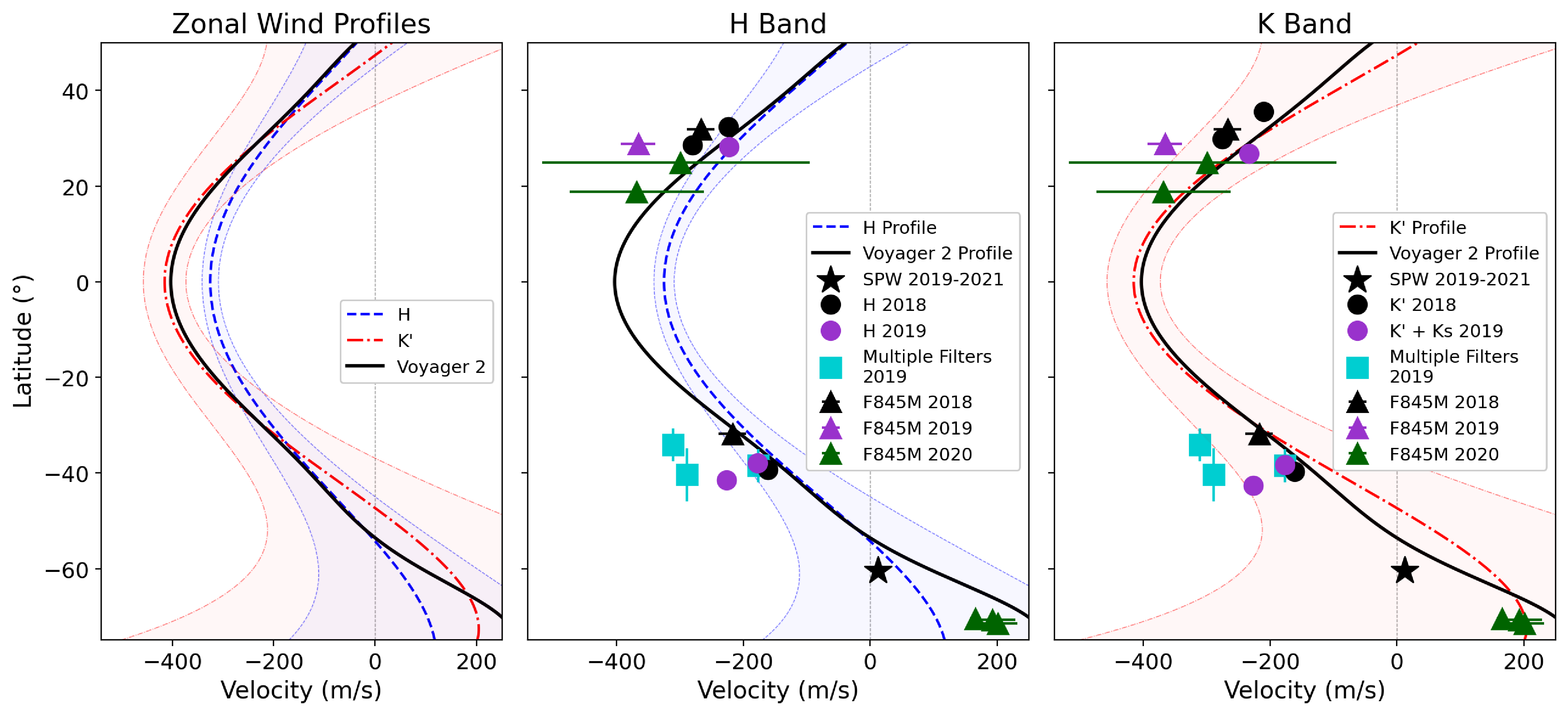}
    \caption{Left: Neptune’s eastward zonal wind profiles and their $1 \sigma$ error as derived by H band data, K’ Band data \citep{tollefson2018} and Voyager 2 data \citep{sromovsky1993}. All latitudes shown are planetographic latitudes. Center: Eastward drift rates of major features compared to the H band and Voyager 2 zonal wind profiles \citep{tollefson2018,sromovsky1993}. Right: Eastward drift rates of major features compared to the K’ Band \citep{tollefson2018} and Voyager 2 zonal wind profiles. All points shown include latitudinal errors, and all except for the blue square points labeled “Multiple Filters” include speed errors. The H band drift rates are compared to the H band profile, and the K’ and Ks band drift rates are compared to the K’ band profile. Eastward drift rates from F845M data, features tracked with multiple filters (including amateur astronomer data), and the South Polar Wave (see Section \ref{sec:spw_analysis}) are compared to both the H and K’ band wind profiles.}
    \label{fig:wind_speeds}
\end{figure}

\begin{deluxetable}{ccccccc}
\tablecaption{Tracked Drift Rates and Latitudes of Major Features \label{tab:drift_rates}}
\tablewidth{0pt}
\tablehead{
\colhead{Date(s)} & \colhead{Telescope(s)} & \colhead{Filter(s)} & \colhead{Latitude}
 & \colhead{Drift Rate} & \colhead{Rotational Period}  & \colhead{Recorded Duration} \\ \colhead{(YYYY-MM-DD)} & & & (\degree) & \colhead{$(m/s)$} & \colhead{(h)}
}
\startdata
\makecell{[2018-05-22, 2018-05-25,\\2018-06-11]} & Keck & H & $28.6 \pm 0.1$ & $-279.85 \pm 1.26$  & $18.2820 \pm 0.0134$ & 20 days \\
\makecell{[2018-05-22, 2018-05-25,\\2018-06-11]} & Keck & K' & $29.9 \pm 0.1$ & $-275.47 \pm 1.53$ & $18.2745 \pm 0.0161$ & 20 days \\
\makecell{[2018-05-23, 2018-05-26]} & Keck & H & $32.4 \pm 0.2$ & $-224.06 \pm 7.05$ & $17.8784 \pm 0.0657$ & 3 days\\
\makecell{[2018-05-23, 2018-05-26]} & Keck & K' & $35.6 \pm 0.2$ & $-210.47 \pm 6.80$ & $17.8303 \pm 0.0667$ & 3 days \\
\makecell{[2018-05-22, 2018-05-23,\\2018-05-25]} & Keck & H & $-39.31 \pm 0.02$ & $-161.14 \pm 6.32$ & $17.4658 \pm 0.0583$ & 3 days \\
\makecell{[2018-05-22, 2018-05-23,\\2018-05-25]} & Keck & K' & $-39.77 \pm 0.02$ & $-160.88 \pm 5.46$ & $17.4732 \pm 0.0508$ & 3 days \\
\makecell{[2019-06-15, 2019-06-17]} & Keck & H & $28.27 \pm 0.03$ & $-222.80 \pm 13.8$ & $17.787 \pm 0.116$ & 2 days \\
\makecell{[2019-06-15, 2019-06-17]} & Keck & K' & $26.82 \pm 0.04$ & $-233.23 \pm 11.0$ & $17.849 \pm 0.092$ & 2 days \\
\makecell{[2019-06-15, 2019-06-17,\\2019-07-04]} & Keck & H & $-37.93 \pm 0.02$ & $-177.74 \pm 0.36$ & $17.5834\pm .0037$ & 19 days \\
\makecell{[2019-06-15, 2019-06-17,\\2019-07-04]} & Keck & K' & $-38.29 \pm 0.03$ & $-176.25 \pm 1.13$ & $17.5816 \pm 0.0109$ & 19 days \\
\makecell{[2019-07-22, 2019-09-04,\\2019-09-05, 2019-09-11,\\ 2019-09-11, 2019-09-12,\\2019-09-25, 2019-09-29,\\ 2019-09-28, 2019-09-29,\\2019-10-03, 2019-10-10,\\2019-10-12]} & \makecell{PCam, GTC,\\Keck, HST,\\A. Data} &  \makecell{RG1000, Z, H\\FQ727N, FQ619N,\\ F845M, F763M,\\ Red, R600} & $-38.5 \pm 3.5$ & $-176.81$ & $17.60$ & 83 days \\
\makecell{[2019-07-19, 2019-07-21,\\2019-10-08, 2019-10-09,\\ 2019-10-10, 2019-10-12]} & PCam, A. Data & RG1000, --- & $-40.4 \pm 5.5$ & $-289.2$ & $18.78$ & 84 days \\
\makecell{[2019-09-05, 2019-09-08,\\2019-09-10, 2019-09-11,\\ 2019-09-28, 2019-09-29,\\2019-10-10]} & \makecell{GTC, Keck,\\A. Data, HST,\\Gemini} & \makecell{Z, H, FQ619N,\\ F845M, FQ727N,\\ F763M, IR785,\\R+IR, R600} & $-34.1 \pm 3.4$ & $-311.12$ & $18.75$ & 46 days \\
\makecell{[2019-09-10, 2019-09-11,\\ 2019-09-14, 2019-11-04]} & Keck, Lick & H & $-41.44 \pm 0.05$ & $-226.48 \pm 0.40$ & $18.1543 \pm .0059$ & 55 days \\
\makecell{[2019-09-10, 2019-09-11,\\ 2019-09-14, 2019-11-04]} & Keck, Lick & K', Ks & $-42.7 \pm 0.1$ & $-226.12 \pm 2.61$ & $18.197 \pm 0.031$ & 55 days \\
\makecell{[2018-09-09, 2018-09-10]} & HST & F845M & $-31.8 \pm 0.6$ & $-217.27 \pm 21.9$ & $17.81 \pm 0.20$ & 19 hrs \\
\makecell{[2018-11-05, 2018-11-06]} & HST & F845M & $31.9 \pm 0.1$ & $-267.17 \pm 21.8$ & $18.05 \pm 0.20$ & 22 hrs \\
\makecell{[2019-09-28, 2019-09-29]} & HST & F845M & $28.8 \pm 0.2$ & $-365.71 \pm 27.5$ & $19.08 \pm 0.27$ & 19 hrs\\
\makecell{[2020-01-07, 2020-01-08]} & HST & F845M & $-71.45 \pm 0.02$ & $200.91 \pm 30.2$ & $13.04 \pm 0.39$ & 4 hrs\\
\makecell{[2020-01-08]} & HST & F845M & $24.96 \pm 0.08$ & $-299.00 \pm 211$ & $18.37 \pm 2.02$ & 40 min\\
\makecell{[2020-06-23]} & HST & F845M & $18.8 \pm 0.3$ & $-368.48 \pm 106$ & $18.84 \pm 0.98$ & 4 hrs \\
\makecell{[2020-06-23]} & HST & F845M & $-70.79 \pm 0.02$ & $192.73 \pm 35.8$ & $13.22 \pm 0.46$ & 4 hrs \\
\makecell{[2020-12-13]} & HST & F845M & $-70.52 \pm 0.04$ & $165.39 \pm 7.77$ & $13.59 \pm 0.10$ & 7 hrs \\
\enddata
\tablecomments{A list of all eastward drift rates of prominent features and their associated latitudes, observed dates, telescopes, associated filters, rotational periods, and recorded durations. A. data stands for Amateur Astronomer data. These data were contributed by Richard Christensen, Marc Delcroix, Darryl Millika, Roberto Sedrani, John Sussenbach, and Anthony Wesley.}
\end{deluxetable}

We implemented a Markov Chain Monte Carlo (MCMC) method in Python using the package “emcee” \citep{foreman-mackey2013emcee}. The two variables determined during the MCMC process were drift rate $(w)$ in degrees per day and initial longitude ($L0$) in degrees. We used the likelihood function:
$$\ln p(L|t, \sigma, w) = -\frac{1}{2} \Sigma_{n} \left[ \frac{(L_{n} - L(w, t, L_0, t_0))^{2}}{s_{n}^{2} } + \ln{(s_{n}^{2})} \right]$$
where $s_{n}^2 = (2\sigma_{n})^2$ is the longitude variance. Note that we double the feature's longitudinal 1-$\sigma$ error in our calculations to use a more realistic uncertainty. Ten independent MCMC chains were used to determine every variable, and 40,000+ steps were used for each chain. The final value of the drift rate $(w)$ was determined by selecting the 50th percentile of the values determined by the MCMC. The upper and lower errors were determined by taking the difference between the central value and the 84th and 16th percentiles, respectively. The Python code used in this process can be found in the emcee package documentation\footnote{https://emcee.readthedocs.io/en/stable/tutorials/line/}.

The resulting drift rates for prominent features were converted from degrees per day to meters per second using a rotation period of 16.11 hours, as derived by \cite{warwick1989}. The latitudinal position of a feature across multiple $(N)$ images ($L_f$) was determined by taking a weighted average of the feature's central longitudes in each image $(L_i)$. The central longitudes were weighted by their errors $\sigma_i$, such that a smaller error increased the weighing of that latitude:

\begin{equation}
    L_f = \sum_{i=1}^{N} \frac{L_i}{\sigma_i^2} / \sum_{i=1}^{N} \frac{1}{\sigma_i^2
}
\end{equation}

The final latitudinal error of that feature was calculated by propagating the latitudinal errors found in the individual images:

\begin{equation}
    \frac{1}{\sigma_{L_f}} = \sqrt{ \frac{1}{\sigma_{L_1}^2}+\frac{1}{\sigma_{L_2}^2}+\dots+\frac{1}{\sigma_{L_i}^2}+\dots+\frac{1}{\sigma_{L_N}^2}}
\end{equation}

These values are plotted against the canonical wind profile determined from Voyager 2 data, and the H and K’ band profiles in Figure \ref{fig:wind_speeds}. There were many more opportunities to calculate drift rates in F845M HST images due to the high cadence of images taken. However, these produced larger error bars due to the small gaps in time between images. The converse is true for the Keck, Lick, and the three multi-filter drift rates --- a larger time coverage resulted in smaller error bars. Details about each drift rate including the latitude, telescopes used, the associated filters, and the range of dates used in determining each speed are summarized in Table \ref{tab:drift_rates}.

There are two cases in which we have a $\sim 2$ month gap between observations of a feature: the features at $-41 \degree$ and $-42 \degree$ during late 2019 seen in H (see Figure \ref{fig:tracking_example}) and K band, respectively. To ensure that we are observing the same feature in both cases, we initially track each feature only considering data from the initial $\sim 5$ days of more frequent observations. We then repeat this tracking after adding the observation taken two months later, and in both instances the MCMC fit returned the same drift rate as the original tracking attempt, but with reduced error bars. This is what we expect if they are in fact the same feature. A large gap in observations can introduce the issue of having multiple drift rates that could potentially fit the data, however the MCMC fit addresses this as well. In both cases, the MCMC fit converges to a a single drift rate that fits the data (see Figure \ref{fig:mcmc_chains_and_speed} for the resulting probability distribution of the H band feature's drift rate), from which we can determine the number of rotations the feature underwent. In cases where we identify a potential feature, but the MCMC fit is unable to converge to a single constant drift rate that describes all the longitude versus time data (typically by fitting all points but one) we conclude that they are not the same feature, and therefore do not report them in this paper.

\section{Analysis of the Bright Ring and the Dark Band \label{sec:bright_ring_dark_band}}

In addition to cloud features identified above, we analyzed two additional features on Neptune that were located near $-60 \degree$ latitude: a bright cloud ring seen in H band, and a dark band prominent in blue wavelengths at around $-65 \degree$ latitude. Both of these features have connections to South Polar Features (SPFs): bright, discrete cloud features with high eastward speeds that appear approximately between $-68 \degree$ and $-75 \degree$ latitude \citep{limayeandsromovsky1991,sromovsky1993}. 
The Bright Ring in H band was sometimes accompanied by SPFs directly to the South (Figure \ref{fig:example_image_grid} and \citealt{chavez2022}), suggesting a possible connection between them. In blue wavelengths, a thick, dark circumpolar band located around $-60 \degree$ is seen on Neptune, known as the South Polar Wave (SPW) \citep{sromovsky2001a,sromovsky2001b}. The SPW remained in a rotational lock with groups of South Polar Features in Voyager 2 and HST data between 1989 and 2010 \citep{sromovsky2001a,sromovsky2001b,karkoschka2011}). 
 
Some of the features tracked with HST data in section \ref{sec:feature_tracking} were SPFs located at latitudes near the Bright Ring and the SPW, motivating further investigation into these two features. We analyze the Bright Ring throughout 2020, the year in which it was most prominent. We then determine the drift rate of the SPW in recent years.

\begin{figure}[h]
    \centering
    \includegraphics[width=0.95\textwidth]{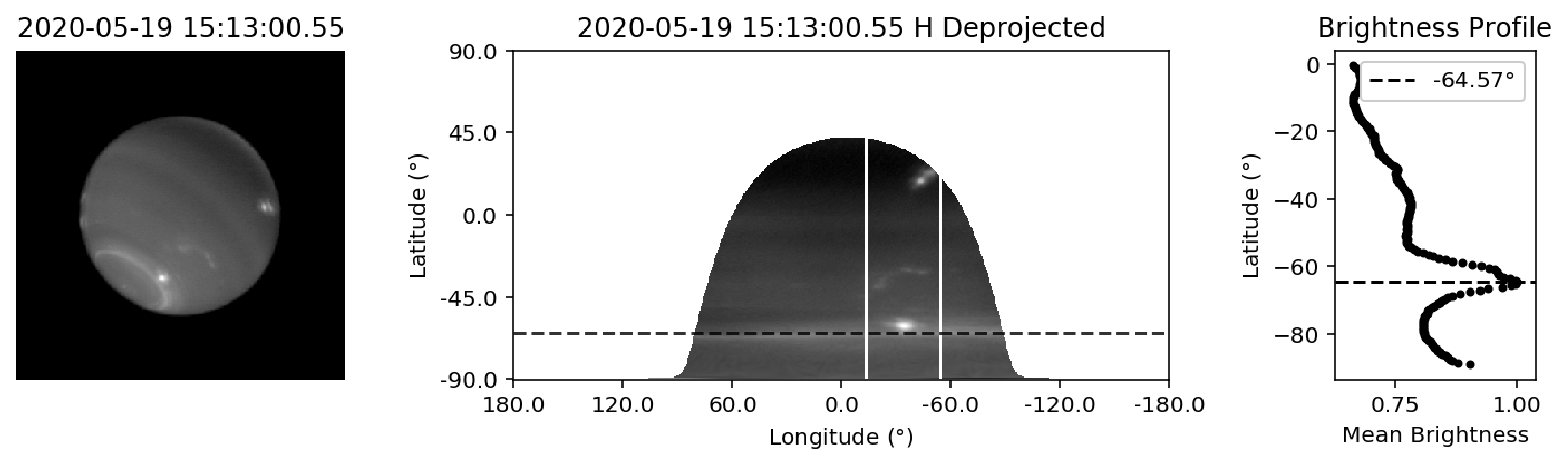}
    \caption{Left: A 2020 H band image. Center: The corresponding projected image with emission angles $(\mu)$ greater than $66 \degree$ removed. Longitudes surrounding the bright feature near the Bright Ring (marked by white vertical lines) were ignored. Right: The normalized latitudinal brightness profile from $90 \degree$ S to the equator made from the central panel. The location of the Bright Ring is overplotted (black dashed line) on the central and right panels).}
    \label{fig:ring_example}
\end{figure}

\subsection{Bright Ring (H band)}

The feature we call the Bright Ring is a narrow, circumferential band of cloud activity at constant latitude located around $-65 \degree$. It was most notable throughout the 2020 H band data as a prominent feature and was sometimes accompanied by SPFs (Figure \ref{fig:ring_example}, \citealt{chavez2022}). The Bright Ring was seen in 2018 and 2019 at similar latitudes \citep{chavez2022}. This established it as a consistent feature on Neptune during recent years at similar latitudes.

\begin{figure}[h]
    \centering
    \includegraphics[width=0.47\textwidth]{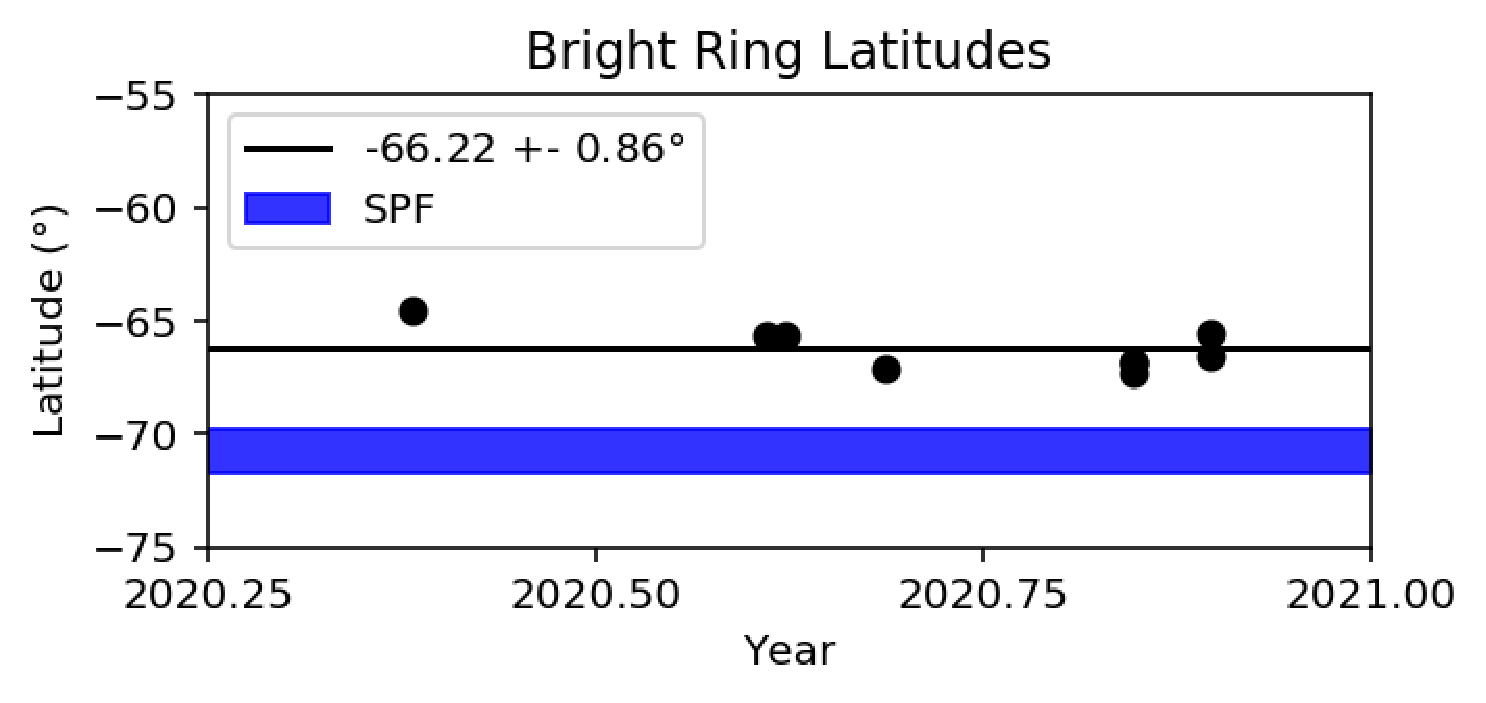}
    \caption{Latitudes of the Bright Ring seen in 2020 with the mean of these latitudes overplotted (black line). The latitude range where South Polar features were tracked (see Table \ref{tab:drift_rates} and Figure \ref{fig:wind_speeds}) is shown (blue shaded area).}
    \label{fig:bright_ring}
\end{figure}

To estimate the central latitude of the Bright Ring, we constructed latitudinal brightness profiles from the 2020 H band data. We began with projected images, first removing edge effects of the projection by excluding all emission angles greater than $66\degree$ ($\mu < 0.4$), as the projection edges can distort the latitudinal position of the ring. We also removed the longitudes where any bright features near $-65\degree$ were present so only the brightness contribution from the Bright Ring was considered. The latitude of the Bright Ring was then calculated by searching for a local maximum in the brightness profile around $-65\degree$. An example of this process is shown in Figure \ref{fig:ring_example}, and the results are shown in Figure \ref{fig:bright_ring}. We determined the ring had an average latitude of $-66.22\degree \pm 0.86 \degree$. The error was determined by taking the standard deviation of the Bright Ring latitudes (see Figure \ref{fig:bright_ring}).

\subsection{Dark Band: South Polar Wave \label{sec:spw}}

In blue wavelengths the SPW, a thick, dark circumpolar wave located around $-60 \degree$ is seen on Neptune. In HST data the SPW has the highest contrast in F467M \citep{sromovsky2001a,sromovsky2001b}, so we used this filter for our analysis. It was originally seen in Voyager 2 data and was most notable for its stability across many years - between 1989 and 2008 the SPW was consistently observed and with an eastward drift rate of $5 \degree/$day \citep{karkoschka2011}. We analyzed the SPW's structure with the goal of tracking its drift rate in recent years (2019 to 2021).

\begin{figure}
    \centering
    \gridline{\fig{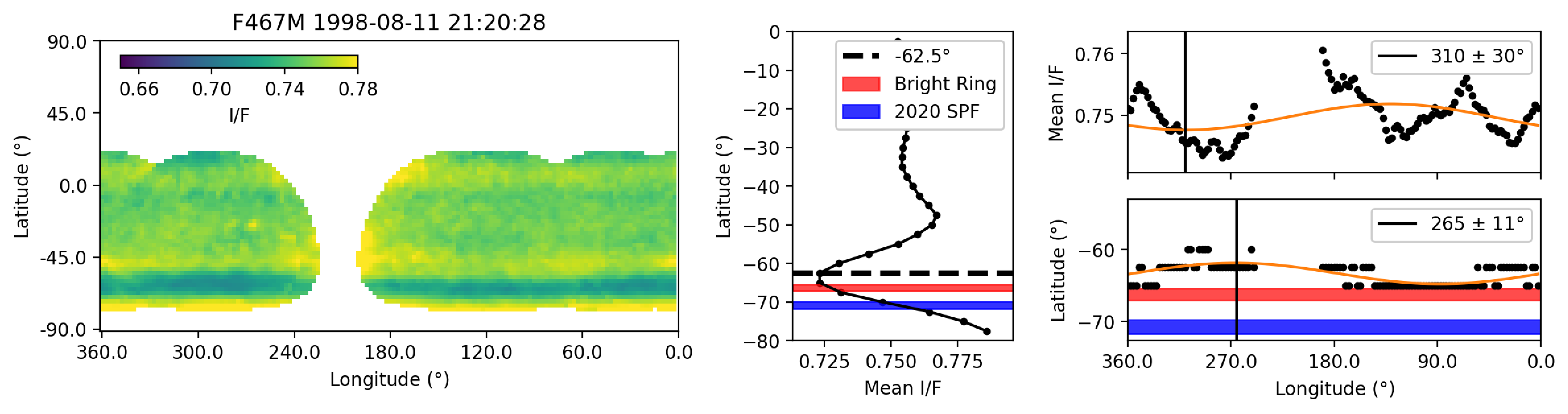}{1.0\textwidth}{}}
    \gridline{\fig{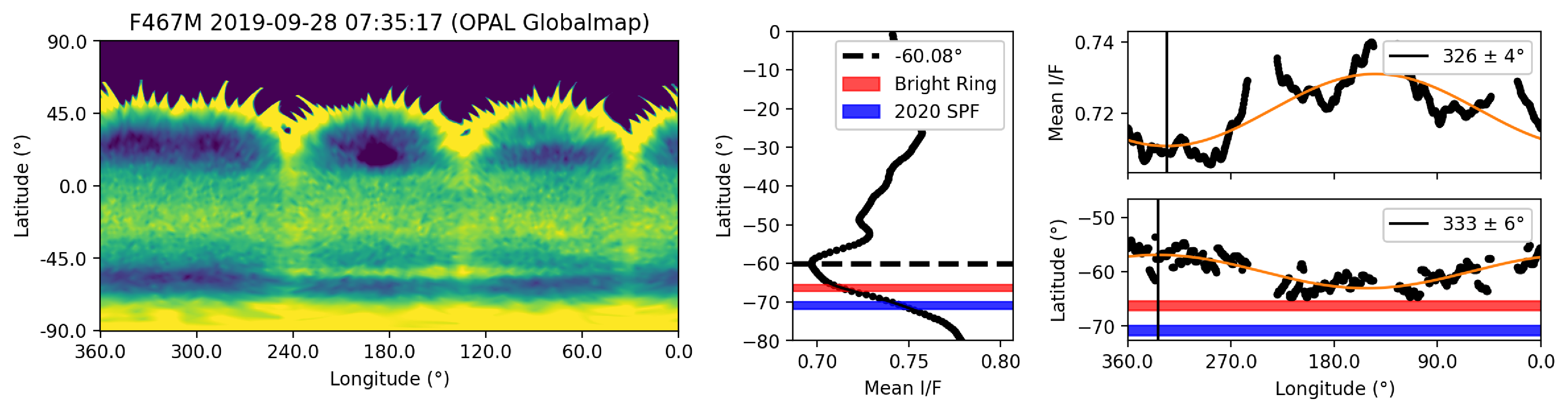}{1.0\textwidth}{}}
    \gridline{\fig{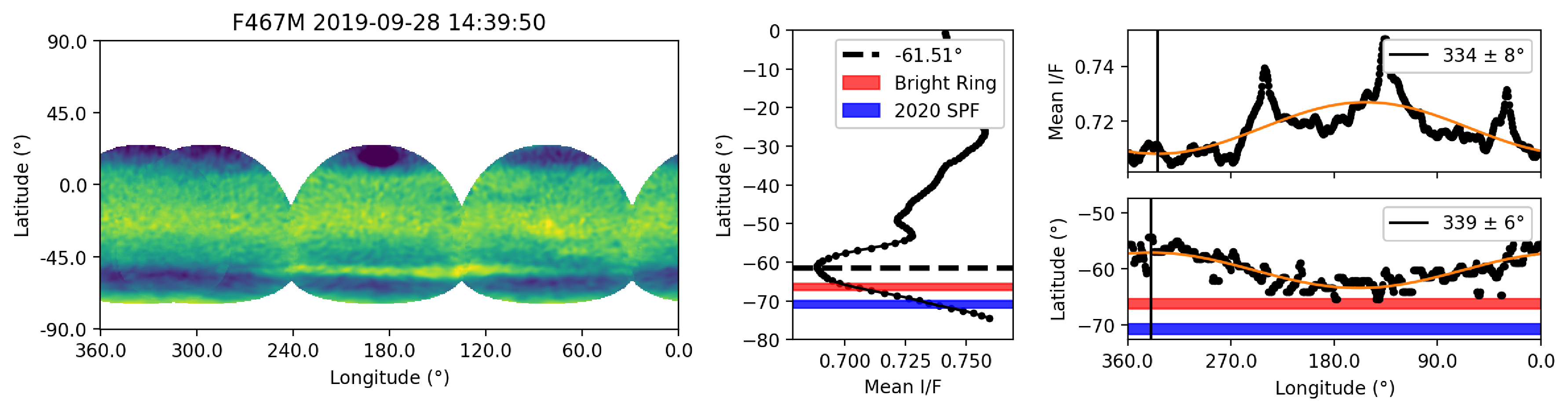}{1.0\textwidth}{}}
    \gridline{\fig{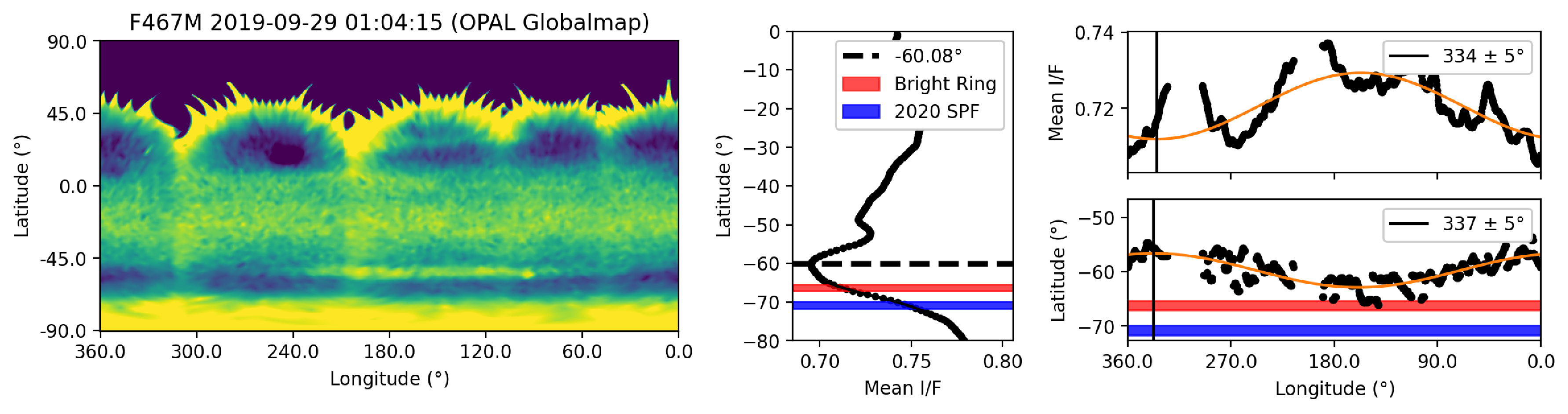}{1.0\textwidth}{}}
\end{figure}
\begin{figure}[t]
    \centering    
    \gridline{\fig{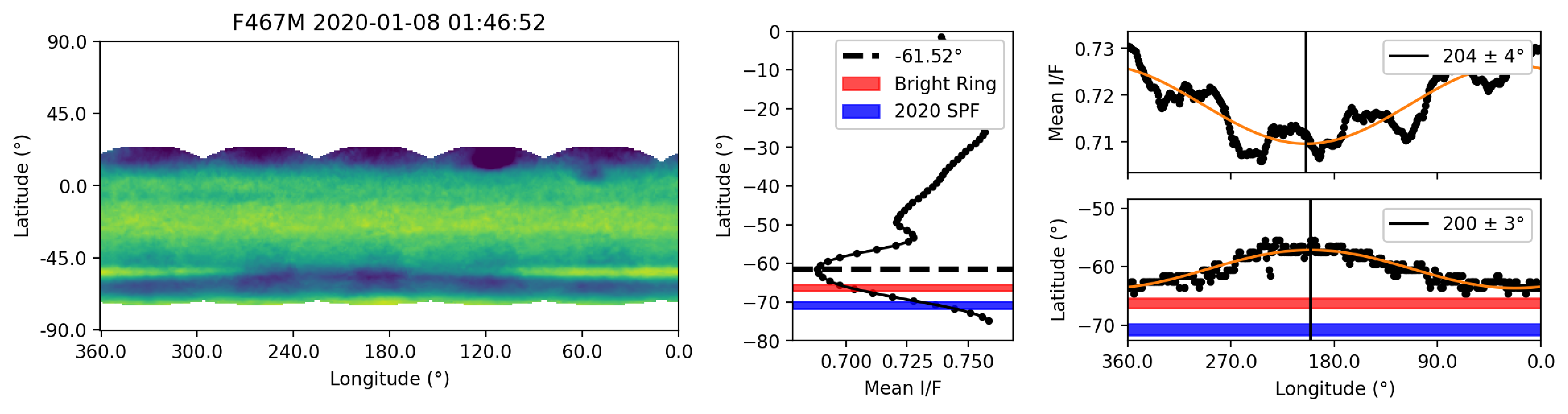}{1.0\textwidth}{}}
    \gridline{\fig{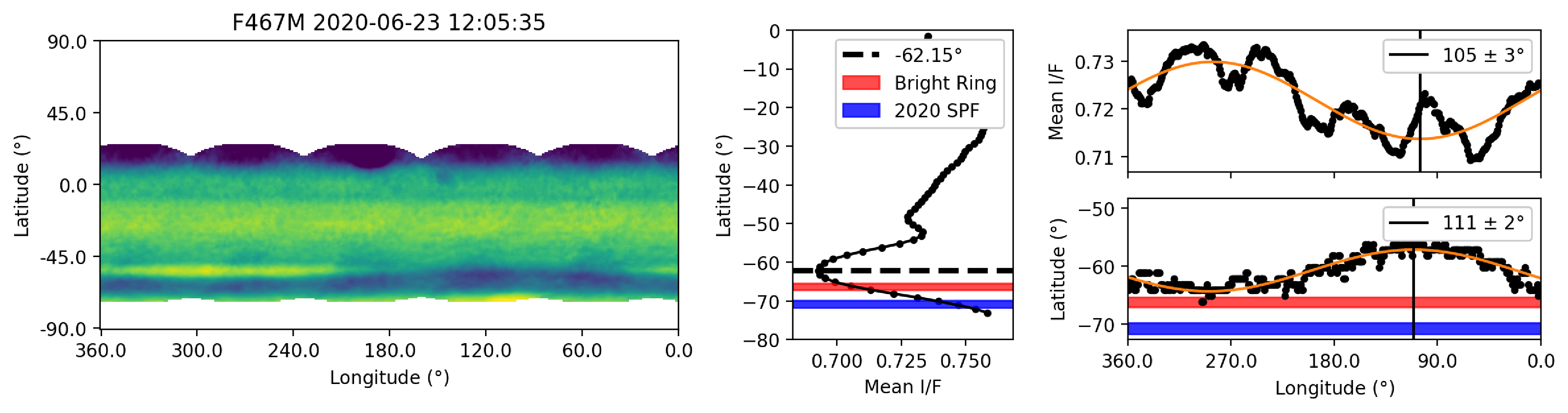}{1.0\textwidth}{}}
    \gridline{\fig{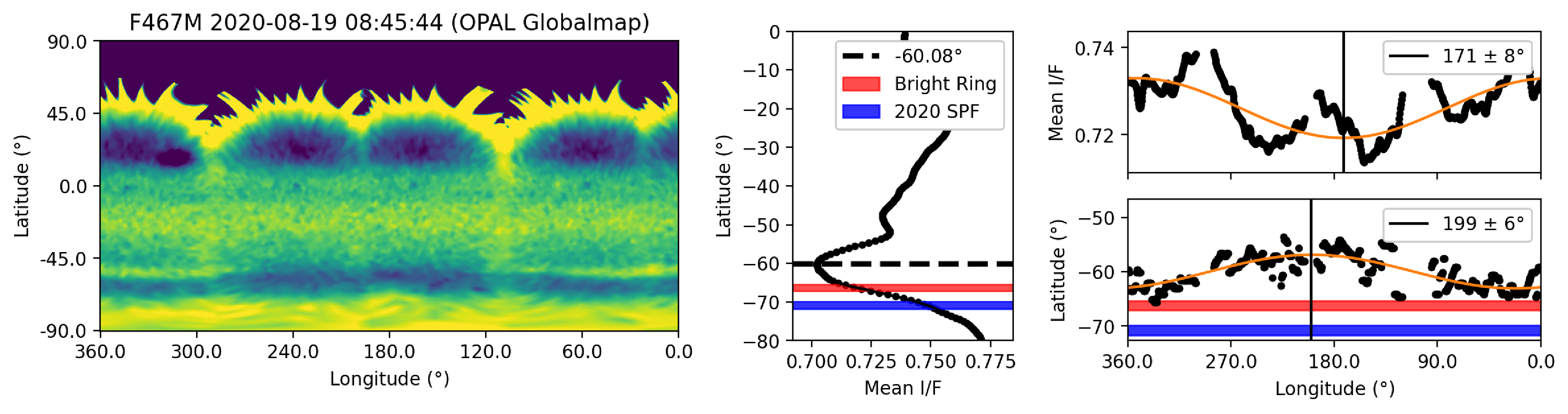}{1.0\textwidth}{}}
    \gridline{\fig{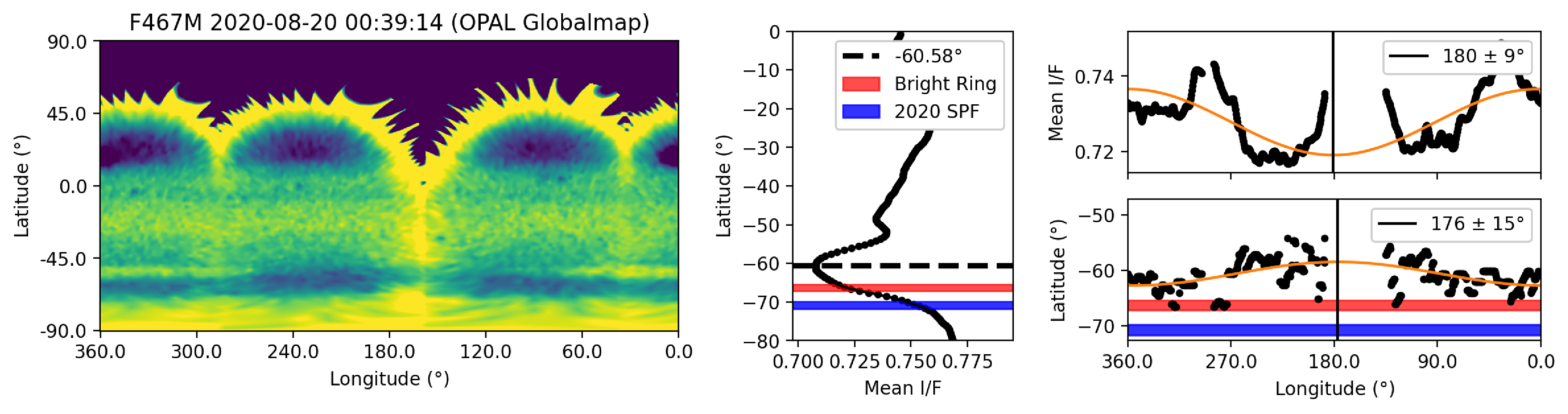}{1.0\textwidth}{}}
\end{figure}
\begin{figure}[t]
    \centering
    \gridline{\fig{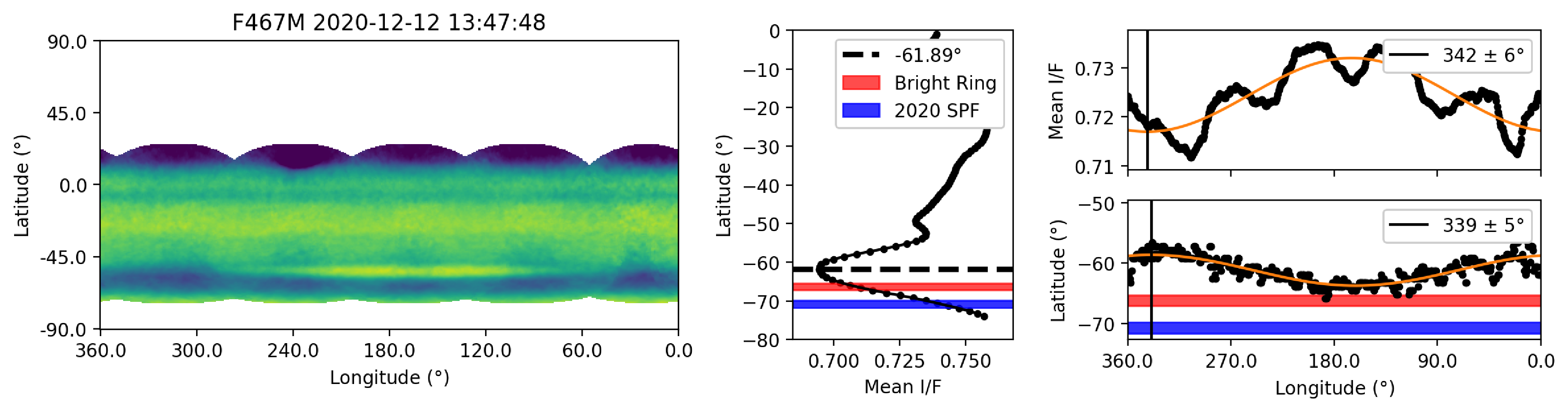}{1.0\textwidth}{}}
    \gridline{\fig{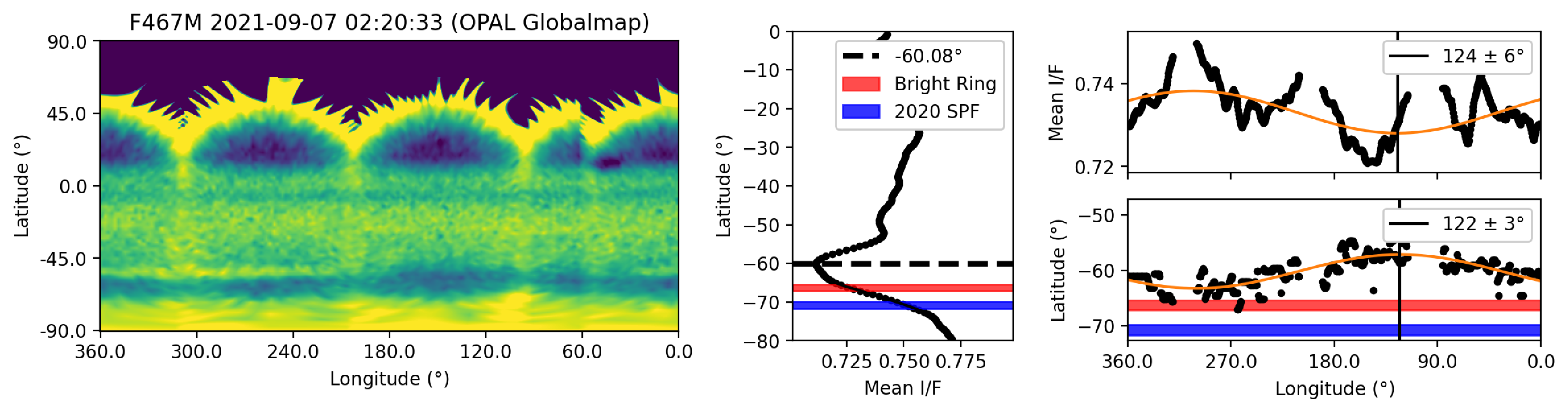}{1.0\textwidth}{}}
    \gridline{\fig{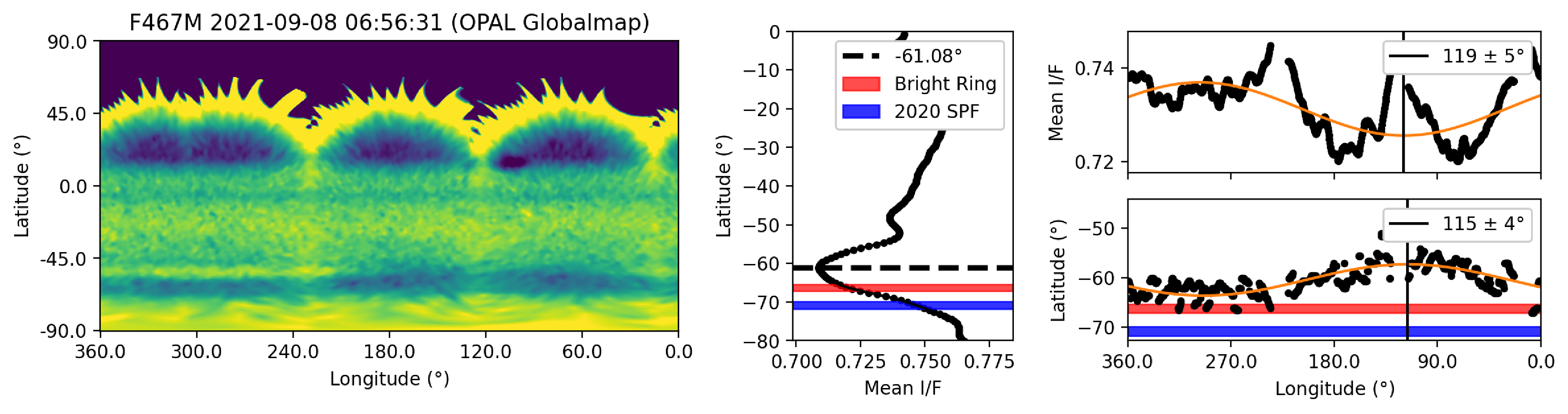}{1.0\textwidth}{}}
    \caption{Left: F467M mosaics used in our analysis. All mosaics are shown with the same image stretch, a color bar is provided in the upper right corner of the August 11, 1998 mosaic. The mosaic image stretch is narrow to bring out the SPW. Center: The average brightness as a function of latitude of these mosaics between the equator and the south pole, shown in units of I/F. Upper Right: The average longitudinal brightness as a function of latitude for the region between $-75 \degree$ and $-45 \degree$ latitude (shown in units of I/F) with a sinusoidal function fit to the data. Longitudes surrounding seams in the OPAL mosaics were removed to minimize non-physical spikes in brightness. The longitude at which the minimum of this function occurs is overplotted (black line). Lower Right: The central latitudinal position of the dark band as a function of longitude with a sinusoidal function fit to the data. The longitude at which the function’s maximum occurs is overplotted (black line). The central and lower right panels include the latitudinal position of the 2020 Bright Ring seen in H band (shaded red area) and the latitudinal positions of SPFs tracked in Section \ref{sec:feature_tracking} (shaded dark blue area).}
    \label{fig:all_mosaics}
\end{figure}

We initially analyzed the SPW during 1998, a year it had previously been observed \citep{sromovsky2001b}, to serve as a point of reference for our recent data. We also used 2019, 2020, and 2021 F467M data to investigate if the SPW was still present and had retained its drift rate in recent years. As the South Polar Wave has a wavelength of $360 \degree$, to see its full extent we stitched multiple images together and constructed mosaics that covered the full $360 \degree$ of Neptune’s longitude. We include two types of mosaics here: some were constructed from using the method described below, while the others were global maps provided by the OPAL program.

\subsubsection{Making F467M Mosaics and Initial Observations of the Wave}

A mosaic was constructed from F467M images taken within a 2-day timespan. As the F467M images were corrected for the effects of limb-darkening (see Section \ref{sec:data}), regions of the disk with emission angles greater than $50 \degree$ were removed to minimize the effects of over-correction. The images were subsequently projected and median averaged to construct the final mosaic. The mean of the exposure start times was used as the mosaic’s time of observation. 

We repeated this process to create one mosaic from 1998 data, another from 2019 data, and three from 2020 data. We also included the mosaiced OPAL global maps. Unlike the mosaics we constructed, the OPAL global maps used interpolation to smooth the seams between images as needed. From these mosaics, we constructed brightness profiles of Neptune's average brightness as a function of latitude, as shown in the central panels of Figure \ref{fig:all_mosaics}. The consistent dip in brightness centered around $-62 \degree$ confirms that the SPW is present in our data (Figure \ref{fig:all_mosaics}). The latitude at which the minimum brightness occurred was used as the central latitude for the dark band in each mosaic. 

We note our initial observations of the dark band, starting with its appearance in 1998. The 1998 data did not complete a full global map; however, a majority of the dark band was still captured. The thickness and latitudinal position of this band were both variable (Figure \ref{fig:all_mosaics}), with thick regions located northwards and thin regions located southwards. The northernmost and southernmost point were separated by approximately $180 \degree$ longitude, confirming a wave structure with a wavelength of $360 \degree$. These characteristics are consistent with previous observations of the South Polar Wave during this time \citep{sromovsky2001b,karkoschka2011}, confirming that the dark band we observed is the same feature as the SPW. 

The SPW was also seen in during 2019, 2020, and 2021 (Figure \ref{fig:all_mosaics}). However, notable changes in structure can be seen in comparison to its 1998 appearance. At longitudes where the band is thin, a corresponding bright band just to the north of the SPW is visible. The thin, bright band and the SPW appear to be complementary, as one is thickest where the other is thinnest and vice-versa. This thin, bright band created a local maximum in the latitudinal brightness profiles centered at $-53 \degree$ latitude (Figure \ref{fig:all_mosaics}), and persisted between 2019 and 2021. 

\subsubsection{Determining Wave Structure and Thickness \label{sec:spw_analysis}}

We created a longitudinal brightness profile of the wave to measure how its thickness varied with longitude (see Figure \ref{fig:all_mosaics}). Longitudes where the wave was thick should result in a lower average brightness, and longitudes where the wave was thin should result in a higher average brightness. 

We isolated the region between $-75 \degree$ and $-45 \degree$ latitude in each mosaic, and calculated the average brightness of this region as a function of longitude. The limb-darkening corrections were less accurate towards the edge of Neptune's disk. This effect combined with the timing of the OPAL data caused bright seams to form in the mosaics which led to maxima within the longitudinal brightness profile (see Figure \ref{fig:all_mosaics}). In the OPAL global maps, we removed longitudes surrounding seams that contributed especially large spikes in the longitudinal brightness profiles. This ensured that patterns we saw in the brightness profile resulted from the SPW and were not a remnant effect of data processing. We then fit a sinusoidal function with wavelength $360 \degree$(wavenumber one) to this profile using non-linear least squares curve fitting \citep{virtanen2020scipy}. The longitude where the SPW was thickest was determined by the minimum of the sinusoidal fit and the longitudinal error was determined by the 1-$\sigma$ error on the sinusoidal phase as given by the non-linear least squares fitting; the results are shown in the upper right panels of Figure \ref{fig:all_mosaics}.

We also calculated the latitude at which the minimum brightness occurred at every longitude, and fit a wavenumber 1 sinusoidal function to these results, once more using non-linear least squares curve fitting \citep{virtanen2020scipy}. The longitude of the band’s northernmost point was determined by the maximum of this sinusoidal fit, and the longitudinal error was again determined by the 1-$\sigma$ error on the sinusoidal phase as given by the non-linear least squares fitting, shown in Figure \ref{fig:all_mosaics}, bottom right panels.

After quantifying how the thickness and position of the SPW vary with longitude, we determined the approximate longitudes where the band was thickest and the longitude of its northernmost point. In our higher resolution data (2019 onwards), we saw that the thickness and position of the were inversely correlated (see Figure \ref{fig:all_mosaics}). This is consistent with \citet{karkoschka2011}, who found that the thickness of the northernmost region of the SPW was thicker than that of the southernmost region. Therefore, we used both the northernmost longitude and thickest longitude as estimates for the central longitude of the wave in every mosaic. We also overplotted the Bright Ring latitudes from the previous section in Figure \ref{fig:all_mosaics} to highlight possible connections between the Bright Ring and the dark band.
 
\begin{figure}
    \centering
    \includegraphics[width=0.99\textwidth]{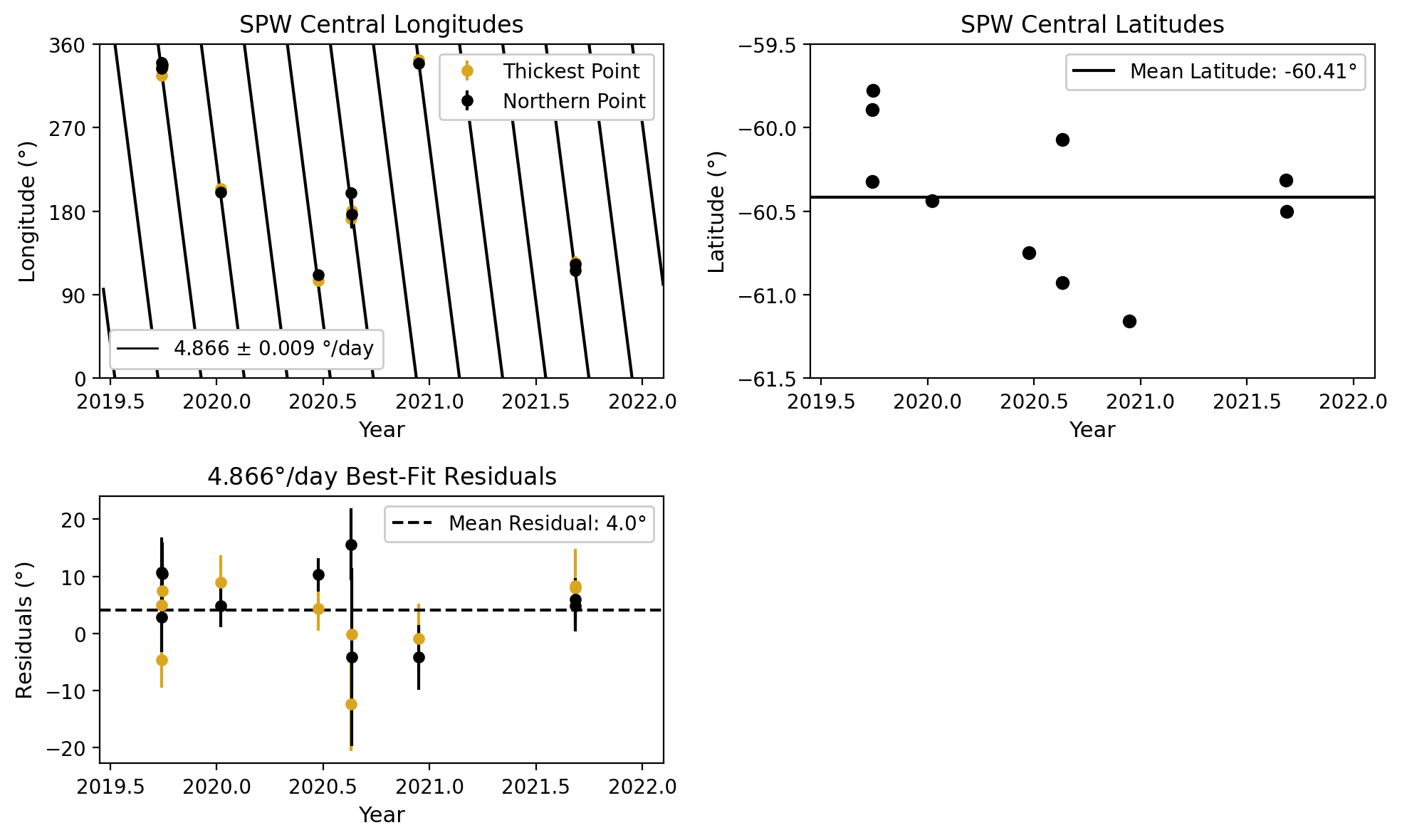}
    \caption{Upper Left: The central longitudes of the SPW (see right panels of Figure \ref{fig:all_mosaics}) versus time in Earth years, with drift rates given in Earth days. The drift rate of $4.866 \pm 0.009 \degree$/day as determined by the MCMC fit is overplotted. Upper Right: The central latitude of the SPW as seen in each mosaic (see central panels of Figure \ref{fig:all_mosaics}). Overplotted is the mean of these central latitudes, which we used as the mean latitude of the SPW between 2019 and 2021. Lower Left: The residuals of the SPW central longitude relative to the resulting MCMC drift rate.}
    \label{fig:spw_drift_rate}
\end{figure}

When considering the SPW in 2019, 2020, and 2021, we used MCMC to identify its drift rate. (see Figure \ref{fig:spw_drift_rate}). We analyzed its drift rate during this time assuming a slow-moving wave at constant speed, therefore we did not consider speeds faster than $10 \degree$/day. With this restriction, we repeated the process outlined in Section \ref{sec:cloud_tracking} with the SPW central longitude versus time as our data. This resulted in an eastward drift rate of $4.866 \pm 0.009 \degree / $day (see Figure \ref{fig:spw_drift_rate}). Averaging the central latitudes of the dark band between 2019 and 2021 resulted in an average central latitude of $-60.39 \degree$ (see Figure \ref{fig:spw_drift_rate}). After assuming the wave moved $4.866 \degree$/day, we determined a speed of $12.02 \pm 0.02$ m/s at the central latitude of $-60.39 \degree$. This speed would be approximately 14 m/s and 10 m/s at the northern and southern ends, respectively, due to the change in latitude.

\section{Results and Discussion \label{sec:discussion}}

\subsection{2018 to 2020 Prominent Features in Near-IR}

We observed that major features within Neptune’s atmosphere clustered in one of 3 regions: the north and south mid-latitudes $(\pm 30 \degree)$ and near the south pole $(\sim -70 \degree)$. This is consistent with previous near-IR observations of Neptune that showed bright mid-latitudinal bands of cloud activity and south polar features \citep{roddier1998,max2003,gibbard2003,martin2012,sromovsky1995,sromovsky2001b,sromovsky2001c}. Features at similar latitudes showed varying drift rates, even when considering features observed within a single filter. While many of the tracked cloud features were consistent with either the Voyager 2 or band-specific wind profiles when taking error bars into account, some were not. Two of the long-lived features tracked with multiple telescopes and amateur astronomer data during 2019 and one northern feature tracked with 2019 F845M data were notably inconsistent with all 3 zonal wind profiles. Similarly, one 2019 H band feature was inconsistent with the H band and Voyager 2 profiles. However, the drift rates of these features might not be representative of the local zonal winds, possibly due to being tied to a dark spot. Analysis from Voyager 2 and HST data show that many of Neptune’s dark spots have bright companion clouds at slightly northern or southern latitudes. These companion clouds co-rotate with the dark spot, and can have drift rates inconsistent with local zonal wind speeds \citep{hammelandlockwood1997,sromovsky2001b,sromovsky2001c}. 

The sporadic temporal coverage of our H band and K' band data meant that features tracked in these filters had lifespans between days and weeks. These features were all observed at mid-latitudes. Various features observed by the higher-cadence F845M data were also located at the mid-latitudes, but we were unable to determine if these features were long-lived. The three longest-lived features that persisted for over a month were observed at southern mid-latitudes. This was consistent with previous long-term tracking results, as between 2013 and 2015 \cite{hueso2017} tracked five long-lived features ($>1$ month) that all existed at southern mid-latitudes. We did not observe long-lived features at northern mid-latitudes, however this could be a result of the observing geometry. Neptune’s south pole is currently tilted towards the Earth, therefore, any long-lived features at northern mid-latitudes had a lower chance of consistently appearing in our observations. With the various prominent features tracked in the northern mid-latitudes, including one feature that persisted for at least 20 days (see Figure \ref{fig:tracked_speeds_keck_lick}), it is highly likely that long-lived features do exist there. A possible explanation for the mid-latitudes containing so many prominent features can be seen in Figure \ref{fig:wind_speeds}. The H and K' band trace different altitudes in Neptune’s atmosphere, with H tracing higher pressures than K’ \citep{tollefson2018}. Different H and K' zonal wind profiles suggest the presence of vertical shear in Neptune’s atmosphere, with the level of shear depending on how drastically the two profiles differ. The mid-latitudes is where all three wind profiles (H band, K' band, and Voyager 2) are roughly equal, suggesting that decreased vertical wind shear could result in higher stability at mid-latitudes.

While 2018 and 2019 showed high levels of cloud activity that provided many opportunities for feature tracking in our H and K' band data, this quickly changed during the start of 2020. The transition between 2019 to 2020 in Neptune’s overall cloud activity was significant. This is described in detail in a companion paper, \cite{chavez2022}. The bright bands of activity and prominent features typical of the mid-latitudes had disappeared. A nearly blank disk with a bright south polar region was typical of the 2020 H and K’ band data and much of the F845M data. No prominent features were tracked in the Keck or Lick data, but the 2020 F845M data were used to track two northern mid-latitude features and three south polar features. The zonal drift rates at northern latitudes were poorly constrained due to limited time coverage, leaving the SPFs as the most precise drift rates calculated. While the drift rates of these SPFs lie within the errors of the H band profile, they align especially well with the K’ band wind profile. As the F845M HST data trace relatively high altitudes in Neptune’s atmosphere, like at K' band (although somewhat lower than K' band), this is consistent with previous findings. 

SPFs that appeared at $-70 \degree$ latitude characterized the prominent cloud activity observed during 2020. The periods of these features (13.1 - 14.6 hr) were consistent with the typical $\sim 13$ hr period of the bright, short-lived SPF features around $-70 \degree$ latitude tracked over a few hours \citep{limayeandsromovsky1991,sromovsky1993,karkoschka2011}. Based upon the available (Voyager and HST) data, SPFs were hypothesized by the latter authors to be composed of many individual cloud features that evolve so rapidly that they cannot be tracked from one rotation to the next. The authors postulate that the features are formed at a particular longitude, move rapidly to the east, and dissipate within hours. This explains their apparent rotation rate of 15.9663 hr \citep{karkoschka2011} when "viewed" over many years, while the rotation rate of individual features is closer to 13 hrs when tracked over several hours. Since we measure the SPF velocity over several hours, we determined the wind speed at which these features move eastward. This speed agrees with the wind profile as measured at K' band, and hence suggests the features to be at high altitudes. This agrees with the interpretation by \cite{smith1989} of the Voyager 2 data, who derived their altitude to be $\sim 50$ km above the cloud deck based upon their shadows cast upon the lower cloud deck. 

Since the 15.9663 hr rotation rate is essentially equal to that of the SPW, \cite{karkoschka2011} suggested these features to be tied to the internal rotation rate of Neptune, and defined a new longitude system, System II, based upon a rotation rate of 15.9663 hr. 

\subsection{Other Features: Bright Ring (H band), South Polar Wave}
Throughout 2020, Neptune was typically seen with a Bright Ring at $-66 \degree $ in H band that is often accompanied by SPFs. From Figure \ref{fig:ring_example}, the Bright Ring was roughly 25\% brighter than the background and had a latitudinal width at half-maximum of about 10 degrees latitude. This Bright Ring was notable due to its prominence during a time where mid-latitudinal cloud activity lowered dramatically, to the point where the mid-latitude cloud bands had disappeared almost entirely. It was seen in H band data since 2018 and its consistent presence (along with its accompanying SPFs) during the 2020 transitional period demonstrate that the south pole’s cloud activity was not suppressed like that at mid-latitudes. 

From our analysis of the dark band in F467M data we saw a clear wave structure in the dark band’s thickness and position in 1998 and subsequently 2019, 2020, and 2021, and we conclude that the South Polar Wave was still present between 2019 to 2021. We also note that the location of the Bright Ring seen in H band typically marked the southern edge of the SPW in recent years (see Figure \ref{fig:all_mosaics}), which suggests that there may be a connection between the two features. 

While the SPW has persisted since the Voyager era, we now observe changes in its structure through the addition of a complementary bright band at $-53 \degree$ latitude. This bright band was only seen within the F467M data. It extended approximately $180 \degree$ across, centered about the southernmost region of the SPW. It had a maximum thickness of approximately $5 \degree$, and its contrast against the surrounding background ranged from tenths of a percent to a few percent (see center panels of Figure \ref{fig:all_mosaics}). Regarding the SPW, we also observe changes in its drift rate in comparison to past observations; our measurement of a $4.866 \pm 0.009 \degree$/day eastward drift rate for the SPW is slightly faster than the $4.828 \pm 0.002 \degree$/day eastward drift rate reported between 1989 and 2010 \citep{karkoschka2011}. However, this change in speed is still consistent with \cite{karkoschka2011}'s result, as the author reported that the SPW had non-zero eastward acceleration of $0.03 \pm 0.02$ m/s per decade. This predicts that the wave's eastward speed would be approximately $4.858 \pm 0.022$ m/s during the time of our measurement a decade later. Our drift rate lies well within this uncertainty and therefore is consistent with \cite{karkoschka2011}'s result. Having retained a nearly constant speed across three decades, South Polar Wave remains as one of Neptune's most stable features.

\section{Conclusions \label{sec:conclusions}}

Using near-infrared data from the Hubble Space Telescope, Keck Observatory, and Lick Observatory supplemented by amateur astronomer data sets, we tracked the zonal drift rates of various prominent cloud features between 2018 and 2020. We also tracked the zonal drift rate of the South Polar Wave using F467M data from 2019 to 2021.

Our results are summarized as follows:
\begin{enumerate}
    \item Between 2018 and 2020, prominent cloud features were tracked only at mid-latitudes and in the south polar region. While features at other latitudes were present, only features at these regions were tracked. The southern mid-latitudes hosted multiple long-lived features ($\geq 1$ month) and was notable as a region of high cloud stability. 
    \item While we were unable to track similarly long-lived features at northern mid-latitudes due to Neptune's viewing geometry partially obstructing the view of its northern hemisphere, one northern feature persisted for at least 20 days, suggesting that the northern mid-latitudes can host long-lived features as well. Consistent observations in the future, ideally with the combination of multiple telescopes to maximize temporal data coverage, are necessary to further explore this possibility and to track the drift rates of these features.
    \item Late 2019/Early 2020 marked a significant decrease in cloud activity characterized by the near-absence of mid-latitude clouds. The South Polar cloud activity remained unaffected and continued to host prominent features throughout 2020. These features were consistent with the K' band zonal wind profile. This change in cloud activity is explored further in \cite{chavez2022}. The prominent features tracked during this period included two northern mid-latitude features with poorly constrained drift rates, and three South Polar Features near $-70 \degree$ latitude. The latter speeds were consistent with those of previously observed bright, short-lived ($< 1$ day) SPF features observed over several hours.
    \item The South Polar Wave seen at blue wavelengths initially observed in Voyager 2 and 1990s HST data is still present in recent years. It has gained a complementary bright band to the North and has slowed by a factor of 6 compared to its appearance between Voyager 2 and 2010. It also may have connections to the Bright Ring at $-66 \degree$ seen in 2020 H band data, as the ring was located at the southern end of the SPW. The SPW drift rate of $4.866 \pm 0.009$ and the corresponding period of $15.9646 \pm 0.0003$ hr is consistent with \citet{karkoschka2011}'s result, which predict a drift rate of $4.858 \pm 0.022$ m/s during the time of our observations.
\end{enumerate}

Infrequent data limited the feature tracking we could accomplish. Frequent and consistent observations in the future, perhaps with the combination of multiple telescopes, are necessary to document the evolution of prominent features on Neptune. However, there is an exciting solution that would address many of the issues we encountered within our data. A Neptune orbiter mission would create opportunities to track features throughout the northern hemisphere, perhaps even at Neptune's North Pole, for extended periods of time.

\section*{Acknowledgments}

Thank you to the two anonymous referees whose comments helped improve the manuscript tremendously.

This work has been supported by the National Science Foundation, NSF Grant AST-1615004 to UC Berkeley.

Many of the images were obtained with the W. M. Keck Observatory, which is operated as a scientific partnership among the California Institute of Technology, the University of California and the National Aeronautics and Space Administration. The Observatory was made possible by the generous financial support of the W. M. Keck Foundation.

The authors wish to recognize and acknowledge the very significant cultural role and reverence that the summit of Maunakea has always had within the indigenous Hawaiian community. We are most fortunate to have the opportunity to conduct observations from this mountain.

We further made use of data obtained with the NASA/ESA Hubble Space Telescope, obtained from the data archive at the Space Telescope Science Institute. 

This work used data acquired from the NASA/ESA HST Space Telescope, associated with OPAL program (PI: Simon, GO13937), and archived by the Space Telescope Science Institute, which is operated by the Association of Universities for Research in Astronomy, Inc., under NASA contract NAS 5-26555. All maps are available at \url{http://dx.doi.org/10.17909/T9G593}.

STScI is operated by the Association of Universities for Research in Astronomy, Inc. under NASA contract NAS 5–26555

Research at Lick Observatory is partially supported by a generous gift from Google.

We are grateful to amateur astronomers performing Neptune observations and submitting them to databases such as PVOL. RH, ASL and JFR were supported by been supported by Grant PID2019-109467GB-I00 funded by MCIN/AEI/10.13039/501100011033/ and by Grupos Gobierno Vasco IT1366-19. Hipercam observations were obtained at Gran Telescopio Canarias (GTC), installed at the Spanish Observatorio del Roque de los Muchachos of the Instituto de Astrofísica de Canarias, on the island of La Palma. Planetcam observations were collected at the Centro Astronómico Hispano en Andalucía (CAHA) at Calar Alto, proposal 19B-2.2-015, operated jointly by Junta de Andalucía and Consejo Superior de Investigaciones Científicas (IAA-CSIC).

\software{Astropy, astroscrappy, emcee, matplotlib, nirc2\_reduce, numpy, pandas, scikit-image, scipy, WinJupos}

\clearpage

\appendix

\setcounter{table}{0}
\renewcommand{\thetable}{A\arabic{table}}
\setcounter{figure}{0}
\renewcommand{\thefigure}{A\arabic{figure}}

\begin{figure}[t]
    \gridline{\fig{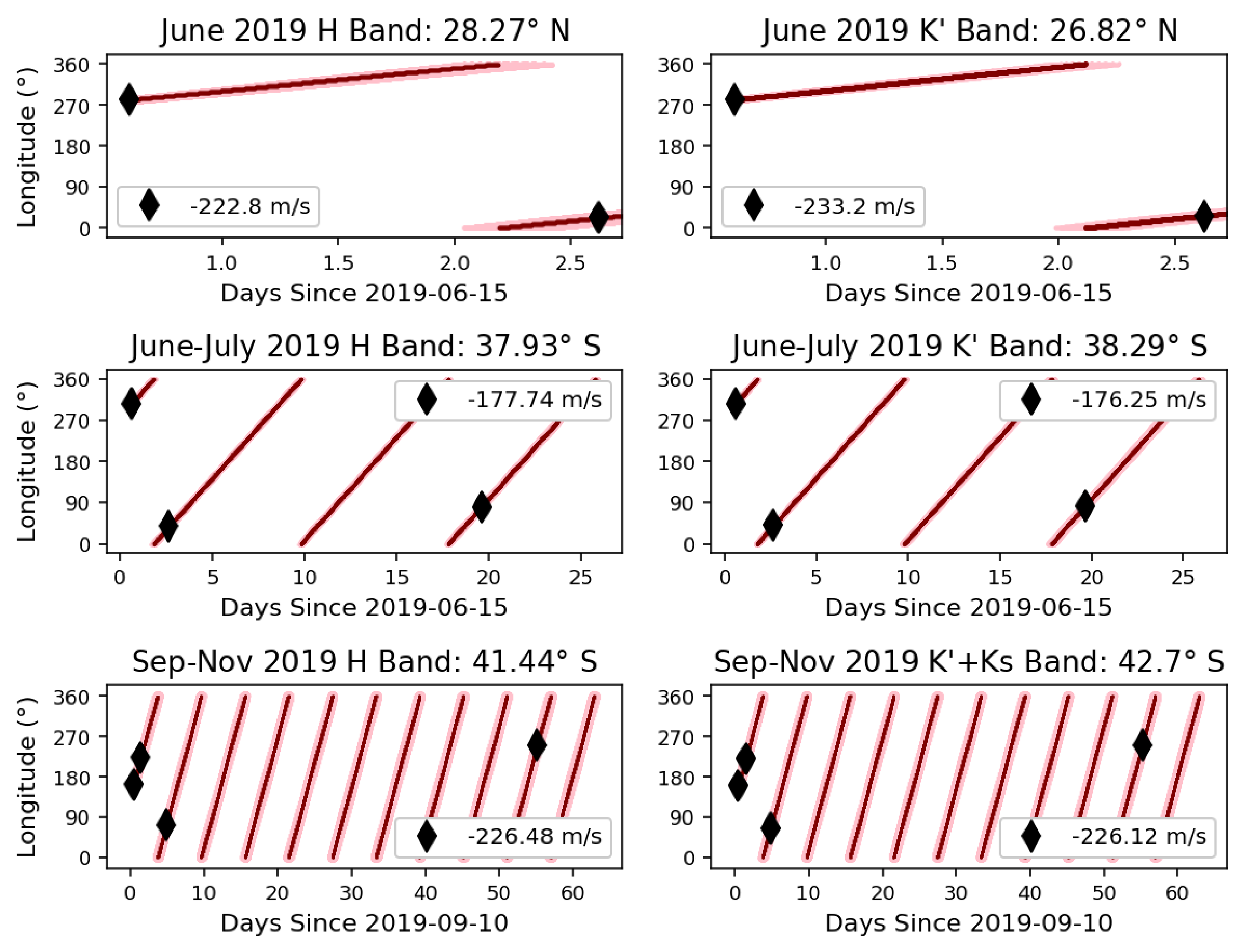}{1\textwidth}{}}
    \caption{Keck and Lick MCMC tracking results of major features with the resulting eastward velocities overplotted.}
    \label{fig:tracked_speeds_keck_lick}
\end{figure}

\begin{figure}[h!]
    \centering
    \includegraphics[width=1.0\textwidth]{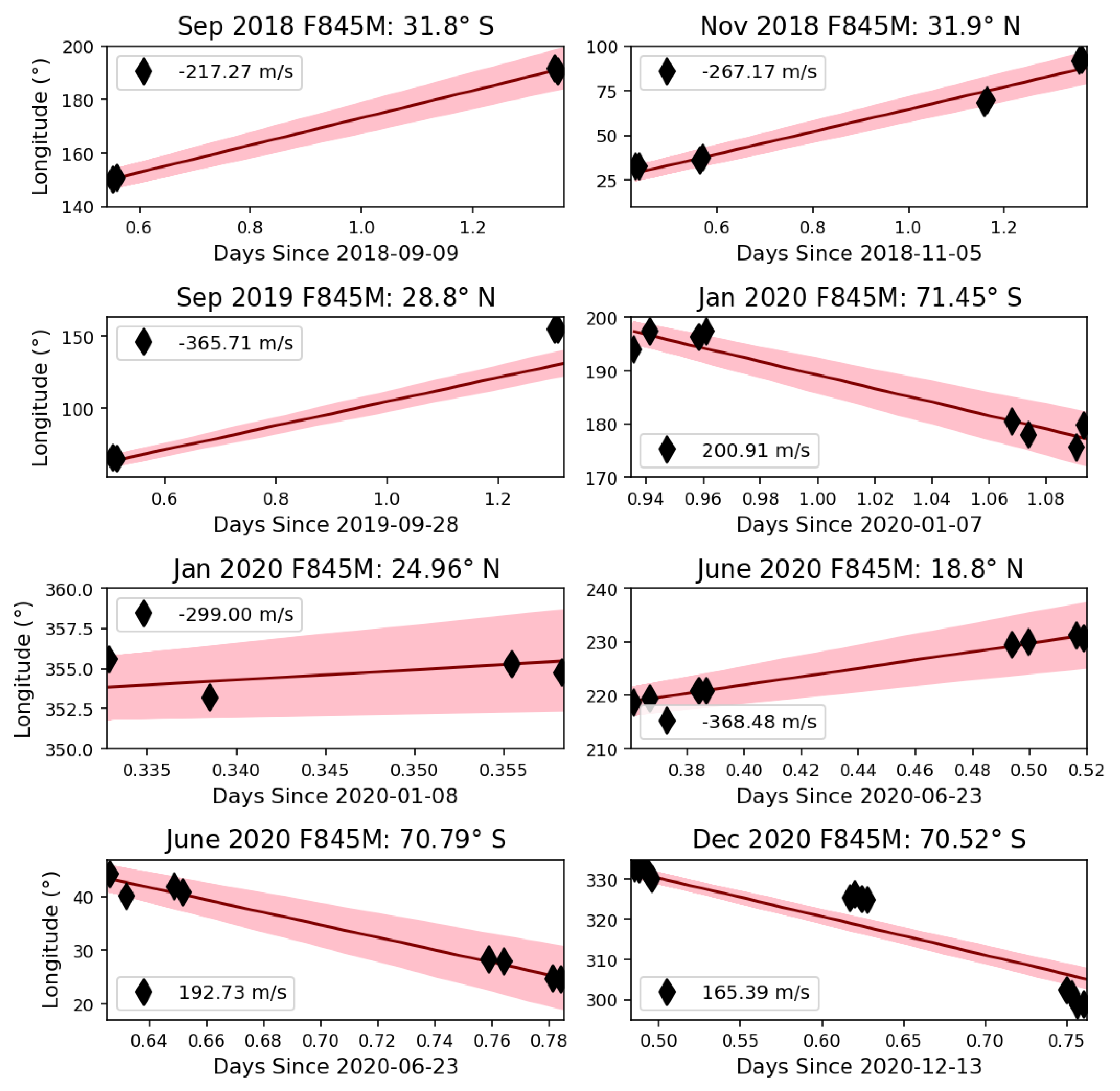}
    \caption{HST MCMC tracking results of major features with the resulting eastward velocities overplotted.}
    \label{fig:tracked_speeds_hst}
\end{figure}

Figures \ref{fig:tracked_speeds_keck_lick} and \ref{fig:tracked_speeds_hst} show MCMC feature tracking results for all the features observed in this paper. Figure \ref{fig:tracked_speeds_keck_lick} shows results from Keck and Lick data while Figure \ref{fig:tracked_speeds_hst} shows results from HST data.
\begin{figure}[h!]
\centering
    \gridline{\fig{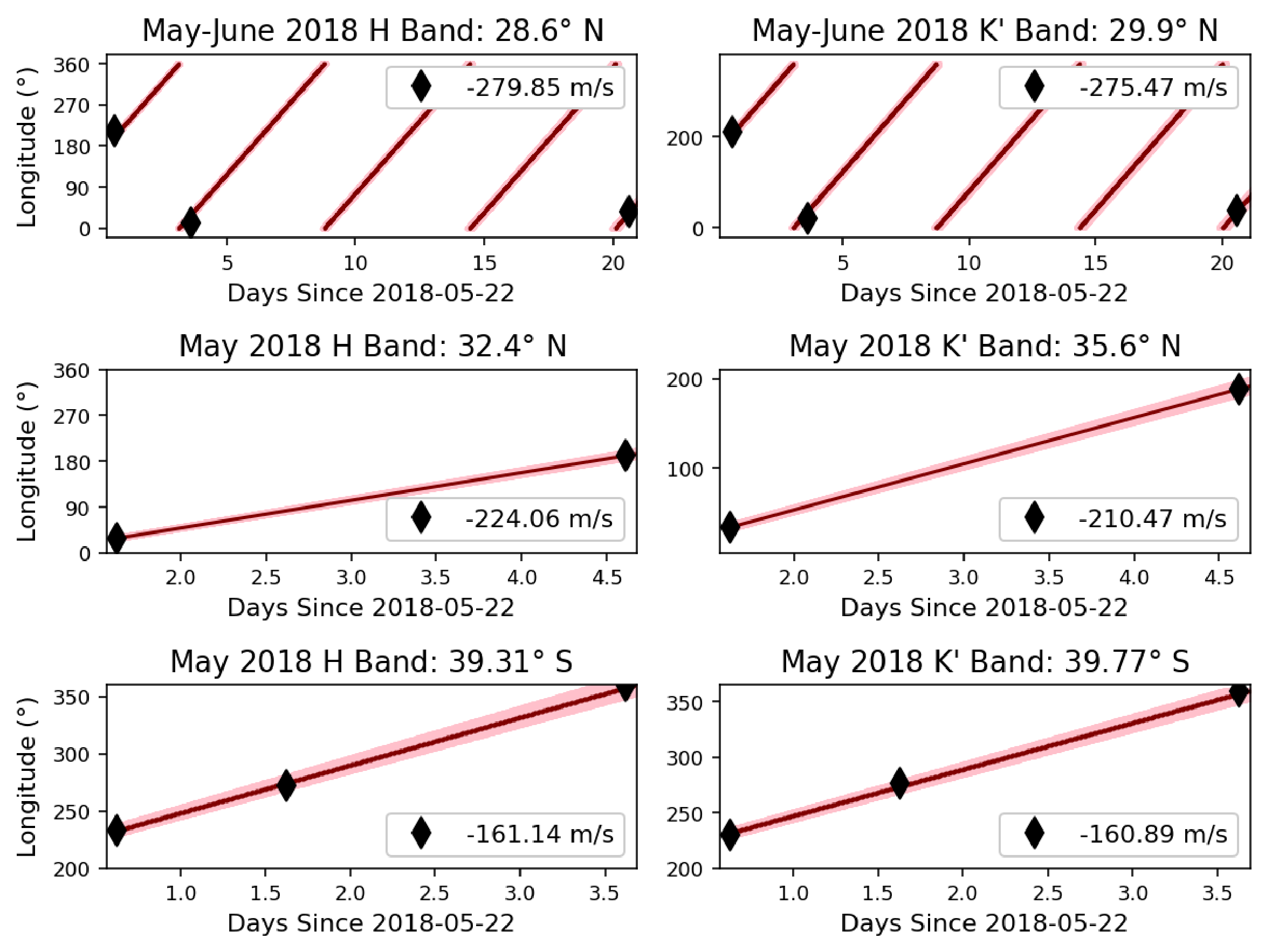}{1\textwidth}{}}
\end{figure}

\clearpage

\begin{figure}[h!]
    \centering
    \includegraphics[width=1.0\textwidth]{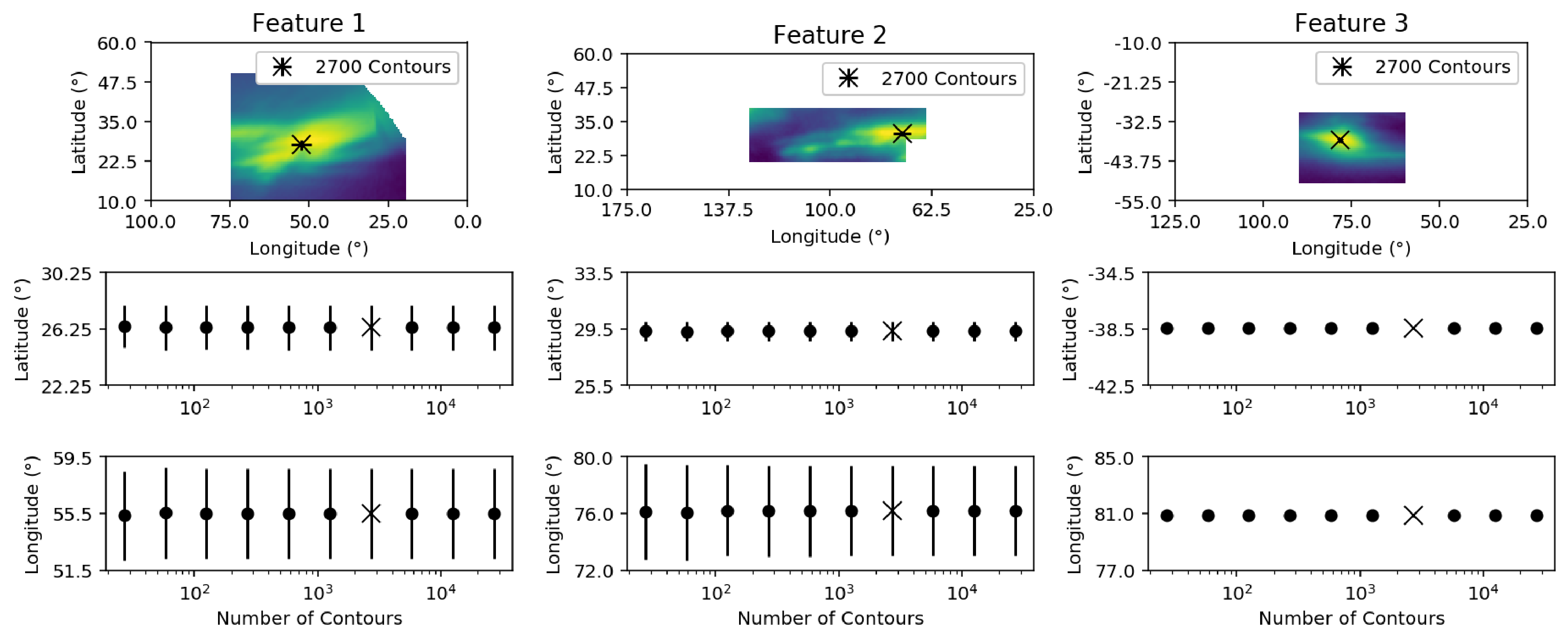}
    \caption{Top Row: Each of the 3 circled features from Figure \ref{fig:tracking_example} are isolated and shown. The central latitude/longitude determined by using 2700 contours between 60\% and 95\% of the feature’s maximum brightness is overplotted. Middle Row: The central latitude of each feature determined by different numbers of contours used. Bottom Row: The central longitude of each feature for different numbers of contours. The value of 2700 contours we used in our analysis is marked with an X in the middle and bottom rows.}
    \label{fig:contours}
\end{figure}

Figure \ref{fig:contours} shows the three features highlighted in Figure \ref{fig:deproj_contours_ex} and their central locations as determined by different numbers of contours. 

Table \ref{tab:near_ir_tracking_images} shows the Keck, Lick, and HST data used for feature tracking and Table \ref{tab:mosaic_images_table} shows the full list of HST data used to construct mosaics in section \ref{sec:spw}.

\startlongtable
\begin{deluxetable}{ccccccc}
\tablecaption{Images Used in Near-IR Feature Tracking \label{tab:near_ir_tracking_images}}
\tablewidth{700pt}
\tabletypesize{\scriptsize}
\tablehead{ \colhead{Telescope} & \colhead{Date (UT)} & \colhead{Time (UT)} & \colhead{Filter} & \colhead{PI/Observers} & 
\colhead{Program} & \colhead{Twilight?}
}
\startdata
Keck & 2018-05-22 & 14:57:05 & H & Alvarez, Mcilroy, Bennett & & Y \\
Keck & 2018-05-22 & 15:02:15 & K' & Alvarez, Mcilroy, Bennett & & Y \\
Keck & 2018-05-23 & 14:55:28 & H & Mcilroy, Ridenour, Alvarez, Bennett & & Y  \\
Keck & 2018-05-23 & 15:00:39 & K' & Mcilroy, Ridenour, Alvarez, Bennett & & Y \\
Keck & 2018-05-25 & 15:00:04 & H & Aycock, Ridenour, Bennett & & Y \\
Keck & 2018-05-25 & 15:05:33 & K' & Aycock, Ridenour, Bennett & & Y \\
Keck & 2018-05-26 & 14:46:03 & H & Aycock, Ridenour, Bennett & & Y \\
Keck & 2018-05-26 & 14:51:14 & K' & Aycock, Ridenour, Bennett & & Y \\
Keck & 2018-06-11 & 14:37:53 & H & Stickel, Alvarez, Hu & & Y \\
Keck & 2018-06-11 & 14:43:07 & K' & Stickel, Alvarez, Hu & & Y \\
HST & 2018-09-09 & 13:15:25 & F845M	& Simon & 15262 & \\	
HST & 2018-09-09 & 13:24:36 & F845M	& Simon & 15262 & \\	
HST & 2018-09-10 & 8:19:48 & F845M & Simon & 15262 & \\		
HST & 2018-09-10 & 8:28:59 & F845M & Simon & 15262 & \\		
HST & 2018-11-05 & 10:20:02 & F845M	& Simon & 15262 & \\
HST & 2018-11-05 & 10:29:13 & F845M	& Simon & 15262 & \\
HST & 2018-11-05 & 13:30:48 & F845M	& Simon & 15262 & \\
HST & 2018-11-05 & 13:39:59 & F845M	& Simon & 15262 & \\
HST & 2018-11-06 & 3:49:37 & F845M & Simon & 15262 & \\		
HST & 2018-11-06 & 3:58:48 & F845M & Simon & 15262 & \\		
HST & 2018-11-06 & 8:35:14 & F845M & Simon & 15262 & \\		
HST & 2018-11-06 & 8:44:25 & F845M & Simon & 15262 & \\		
Keck & 2019-06-15 & 14:28:44 & H & Pelletier, de Kleer & & Y\\
Keck & 2019-06-15 & 14:33:46 & K' & Pelletier, de Kleer & & Y\\
Keck & 2019-06-17 & 14:54:24 & H & Pelletier, Gaidos & & Y \\
Keck & 2019-07-04 & 15:09:32 & H & Team Keck & & Y \\
Keck & 2019-09-10 & 11:25:14 & H & Sromovsky, Fry, de Pater & \\
Keck & 2019-09-10 & 11:30:45 & K' & Sromovsky, Fry, de Pater & \\
Keck & 2019-09-11 & 11:19:01 & H & Sromovsky, Fry, de Pater & \\
Keck & 2019-09-11 & 11:24:21 & K' & Sromovsky, Fry, de Pater & \\
Lick & 2019-09-14 & 8:45:57 & H & Gates, Giacalone, Dressing & & Y\\		
Lick & 2019-09-14 & 9:01:38 & Ks & Gates, Giacalone, Dressing & & Y\\		
HST & 2019-09-28 & 12:10:09 & F845M & Simon & 15502 &  \\
HST & 2019-09-28 & 12:19:18 & F845M & Simon & 15502 &  \\
HST & 2019-09-29 & 7:14:28 & F845M & Simon & 15502 &  \\		
HST & 2019-09-29 & 7:23:37 & F845M & Simon & 15502 &  \\		
Keck & 2019-11-04 & 4:14:21 & H & de Pater, Tollefson & \\		
Keck & 2019-11-04 & 4:21:05 & K' & de Pater, Tollefson & \\		
HST & 2020-01-07 & 22:27:22 & F845M	& Wong & 16057 & \\
HST & 2020-01-07 & 22:35:30 & F845M	& Wong & 16057 & \\
HST & 2020-01-07 & 22:59:00 & F845M & Wong & 16057 & \\		
HST & 2020-01-07 & 23:03:53 & F845M & Wong & 16057 & \\
HST & 2020-01-08 & 1:38:02 & F845M & Wong & 16057 & \\		
HST & 2020-01-08 & 1:46:10 & F845M & Wong & 16057 & \\		
HST & 2020-01-08 & 2:10:31 & F845M & Wong & 16057 & \\		
HST & 2020-01-08 & 2:14:33 & F845M & Wong & 16057 & \\		
HST & 2020-01-08 & 7:59:20 & F845M & Wong & 16057 & \\		
HST & 2020-01-08 & 8:07:28 & F845M & Wong & 16057 & \\		
HST & 2020-01-08 & 8:31:49 & F845M & Wong & 16057 & \\		
HST & 2020-01-08 & 8:35:51 & F845M & Wong & 16057 & \\		
HST & 2020-06-23 & 8:40:29 & F845M & Wong & 16084 & \\		
HST & 2020-06-23 & 8:48:36 & F845M & Wong & 16084 & \\		
HST & 2020-06-23 & 9:12:57 & F845M & Wong & 16084 & \\		
HST & 2020-06-23 & 9:16:59 & F845M & Wong & 16084 & \\		
HST & 2020-06-23 & 11:51:08 & F845M & Wong & 16084 & \\
HST & 2020-06-23 & 11:59:16 & F845M & Wong & 16084 & \\
HST & 2020-06-23 & 12:23:37 & F845M & Wong & 16084 & \\
HST & 2020-06-23 & 12:27:39 & F845M & Wong & 16084 & \\
HST & 2020-06-23 & 15:01:50 & F845M & Wong & 16084 & \\
HST & 2020-06-23 & 15:09:58 & F845M & Wong & 16084 & \\
HST & 2020-06-23 & 15:34:19 & F845M & Wong & 16084 & \\
HST & 2020-06-23 & 15:38:21 & F845M & Wong & 16084 & \\
HST & 2020-06-23 & 18:12:30 & F845M & Wong & 16084 & \\
HST & 2020-06-23 & 18:20:38 & F845M & Wong & 16084 & \\
HST & 2020-06-23 & 18:44:59 & F845M & Wong & 16084 & \\
HST & 2020-06-23 & 18:49:01 & F845M & Wong & 16084 & \\
HST & 2020-12-13 & 11:38:31 & F845M & Wong & 16454 & \\
HST & 2020-12-13 & 11:42:33 & F845M & Wong & 16454 & \\
HST & 2020-12-13 & 11:48:07 & F845M & Wong & 16454 & \\
HST & 2020-12-13 & 11:53:41 & F845M & Wong & 16454 & \\
HST & 2020-12-13 & 14:49:11 & F845M & Wong & 16454 & \\
HST & 2020-12-13 & 14:53:13 & F845M & Wong & 16454 & \\
HST & 2020-12-13 & 14:58:47 & F845M & Wong & 16454 & \\
HST & 2020-12-13 & 15:04:21 & F845M & Wong & 16454 & \\
HST & 2020-12-13 & 17:59:50 & F845M & Wong & 16454 & \\
HST & 2020-12-13 & 18:03:52 & F845M & Wong & 16454 & \\
HST & 2020-12-13 & 18:09:26 & F845M & Wong & 16454 & \\
HST & 2020-12-13 & 18:15:00 & F845M & Wong & 16454 & \\
\enddata
\end{deluxetable}

\startlongtable
\begin{deluxetable}{cccccc}
\tablecaption{HST F467M Images Used to Construct Mosaics\label{tab:mosaic_images_table}}
\tablewidth{700pt}
\tabletypesize{\scriptsize}
\tablehead{ \colhead{Telescope} & \colhead{Date (UT)} & \colhead{Time (UT)} & \colhead{PI} & \colhead{Program}
} 
\startdata
HST & 1998-08-11 & 5:20:13 & Sromovsky & 7324\\
HST & 1998-08-11 & 6:56:13 & Sromovsky & 7324\\
HST & 1998-08-11 & 10:11:13 & Sromovsky & 7324\\
HST & 1998-08-11 & 11:46:13 & Sromovsky & 7324\\
HST & 1998-08-12 & 2:53:13 & Sromovsky & 7324\\
HST & 1998-08-12 & 10:55:13 & Sromovsky & 7324\\
HST & 1998-08-12 & 10:55:13 & Sromovsky & 7324\\
HST & 1998-08-12 & 15:46:13 & Sromovsky & 7324\\
HST & 2019-09-28 & 7:26:08 & Simon & 15502\\
HST & 2019-09-28 & 7:35:17 & Simon & 15502\\
HST & 2019-09-28 & 12:12:13 & Simon & 15502\\
HST & 2019-09-28 & 12:21:22 & Simon & 15502\\
HST & 2019-09-28 & 16:58:18 & Simon & 15502\\
HST & 2019-09-28 & 17:07:27 & Simon & 15502\\
HST & 2019-09-28 & 21:44:23 & Simon & 15502\\
HST & 2019-09-28 & 21:53:32 & Simon & 15502\\
HST & 2020-01-07 & 19:12:16 & Wong & 16057\\
HST & 2020-01-07 & 19:18:42 & Wong & 16057\\
HST & 2020-01-07 & 19:20:24 & Wong & 16057\\
HST & 2020-01-07 & 19:26:50 & Wong & 16057\\
HST & 2020-01-07 & 19:30:32 & Wong & 16057\\
HST & 2020-01-07 & 19:34:14 & Wong & 16057\\
HST & 2020-01-07 & 19:51:11 & Wong & 16057\\
HST & 2020-01-07 & 19:55:13 & Wong & 16057\\
HST & 2020-01-07 & 22:22:56 & Wong & 16057\\
HST & 2020-01-07 & 22:29:22 & Wong & 16057\\
HST & 2020-01-07 & 22:31:04 & Wong & 16057\\
HST & 2020-01-07 & 22:37:30 & Wong & 16057\\
HST & 2020-01-07 & 22:41:12 & Wong & 16057\\
HST & 2020-01-07 & 22:44:54 & Wong & 16057\\
HST & 2020-01-07 & 23:01:51 & Wong & 16057\\
HST & 2020-01-07 & 23:05:53 & Wong & 16057\\
HST & 2020-01-08 & 1:33:36 & Wong & 16057\\
HST & 2020-01-08 & 1:40:02 & Wong & 16057\\
HST & 2020-01-08 & 1:41:44 & Wong & 16057\\
HST & 2020-01-08 & 1:48:10 & Wong & 16057\\
HST & 2020-01-08 & 1:51:52 & Wong & 16057\\
HST & 2020-01-08 & 2:12:31 & Wong & 16057\\
HST & 2020-01-08 & 2:16:33 & Wong & 16057\\
HST & 2020-01-08 & 4:44:16 & Wong & 16057\\
HST & 2020-01-08 & 4:50:42 & Wong & 16057\\
HST & 2020-01-08 & 4:52:24 & Wong & 16057\\
HST & 2020-01-08 & 4:58:50 & Wong & 16057\\
HST & 2020-01-08 & 5:02:32 & Wong & 16057\\
HST & 2020-01-08 & 5:06:14 & Wong & 16057\\
HST & 2020-01-08 & 5:27:13 & Wong & 16057\\
HST & 2020-01-08 & 7:54:54 & Wong & 16057\\
HST & 2020-01-08 & 8:01:20 & Wong & 16057\\
HST & 2020-01-08 & 8:03:02 & Wong & 16057\\
HST & 2020-01-08 & 8:09:28 & Wong & 16057\\
HST & 2020-01-08 & 8:13:10 & Wong & 16057\\
HST & 2020-01-08 & 8:16:52 & Wong & 16057\\
HST & 2020-01-08 & 8:33:49 & Wong & 16057\\
HST & 2020-01-08 & 8:37:51 & Wong & 16057\\
HST & 2020-06-23 & 5:25:21 & Wong & 16084\\
HST & 2020-06-23 & 5:31:47 & Wong & 16084\\
HST & 2020-06-23 & 5:33:29 & Wong & 16084\\
HST & 2020-06-23 & 5:39:55 & Wong & 16084\\
HST & 2020-06-23 & 5:43:37 & Wong & 16084\\
HST & 2020-06-23 & 5:47:19 & Wong & 16084\\
HST & 2020-06-23 & 6:04:16 & Wong & 16084\\
HST & 2020-06-23 & 6:08:18 & Wong & 16084\\
HST & 2020-06-23 & 8:36:02 & Wong & 16084\\
HST & 2020-06-23 & 8:42:28 & Wong & 16084\\
HST & 2020-06-23 & 8:44:10 & Wong & 16084\\
HST & 2020-06-23 & 8:50:36 & Wong & 16084\\
HST & 2020-06-23 & 8:54:18 & Wong & 16084\\
HST & 2020-06-23 & 8:58:00 & Wong & 16084\\
HST & 2020-06-23 & 9:14:57 & Wong & 16084\\
HST & 2020-06-23 & 9:18:59 & Wong & 16084\\
HST & 2020-06-23 & 11:46:42 & Wong & 16084\\
HST & 2020-06-23 & 11:53:08 & Wong & 16084\\
HST & 2020-06-23 & 11:54:50 & Wong & 16084\\
HST & 2020-06-23 & 12:01:16 & Wong & 16084\\
HST & 2020-06-23 & 12:04:58 & Wong & 16084\\
HST & 2020-06-23 & 12:08:40 & Wong & 16084\\
HST & 2020-06-23 & 12:25:37 & Wong & 16084\\
HST & 2020-06-23 & 12:29:39 & Wong & 16084\\
HST & 2020-06-23 & 14:57:24 & Wong & 16084\\
HST & 2020-06-23 & 15:03:50 & Wong & 16084\\
HST & 2020-06-23 & 15:03:50 & Wong & 16084\\
HST & 2020-06-23 & 15:11:58 & Wong & 16084\\
HST & 2020-06-23 & 15:15:40 & Wong & 16084\\
HST & 2020-06-23 & 15:19:22 & Wong & 16084\\
HST & 2020-06-23 & 15:36:19 & Wong & 16084\\
HST & 2020-06-23 & 15:40:21 & Wong & 16084\\
HST & 2020-06-23 & 18:08:04 & Wong & 16084\\
HST & 2020-06-23 & 18:14:30 & Wong & 16084\\
HST & 2020-06-23 & 18:16:12 & Wong & 16084\\
HST & 2020-06-23 & 18:22:38 & Wong & 16084\\
HST & 2020-06-23 & 18:26:20 & Wong & 16084\\
HST & 2020-06-23 & 18:30:02 & Wong & 16084\\
HST & 2020-06-23 & 18:46:59 & Wong & 16084\\
HST & 2020-06-23 & 18:51:01 & Wong & 16084\\
HST & 2020-12-12 & 6:43:43 & Wong & 16454\\
HST & 2020-12-12 & 6:45:25 & Wong & 16454\\
HST & 2020-12-12 & 6:47:07 & Wong & 16454\\
HST & 2020-12-12 & 6:48:49 & Wong & 16454\\
HST & 2020-12-12 & 6:50:31 & Wong & 16454\\
HST & 2020-12-12 & 6:52:13 & Wong & 16454\\
HST & 2020-12-12 & 6:55:41 & Wong & 16454\\
HST & 2020-12-12 & 6:59:09 & Wong & 16454\\
HST & 2020-12-12 & 7:10:41 & Wong & 16454\\
HST & 2020-12-12 & 7:16:15 & Wong & 16454\\
HST & 2020-12-12 & 7:21:49 & Wong & 16454\\
HST & 2020-12-12 & 7:23:31 & Wong & 16454\\
HST & 2020-12-12 & 7:25:13 & Wong & 16454\\
HST & 2020-12-12 & 7:26:55 & Wong & 16454\\
HST & 2020-12-12 & 9:54:23 & Wong & 16454\\
HST & 2020-12-12 & 9:56:05 & Wong & 16454\\
HST & 2020-12-12 & 9:57:47 & Wong & 16454\\
HST & 2020-12-12 & 9:59:29 & Wong & 16454\\
HST & 2020-12-12 & 10:01:11 & Wong & 16454\\
HST & 2020-12-12 & 10:02:53 & Wong & 16454\\
HST & 2020-12-12 & 13:05:03 & Wong & 16454\\
HST & 2020-12-12 & 13:06:45 & Wong & 16454\\
HST & 2020-12-12 & 13:08:27 & Wong & 16454\\
HST & 2020-12-12 & 13:10:09 & Wong & 16454\\
HST & 2020-12-12 & 13:11:51 & Wong & 16454\\
HST & 2020-12-12 & 13:13:33 & Wong & 16454\\
HST & 2020-12-12 & 13:17:01 & Wong & 16454\\
HST & 2020-12-12 & 13:20:29 & Wong & 16454\\
HST & 2020-12-12 & 13:32:01 & Wong & 16454\\
HST & 2020-12-12 & 13:37:35 & Wong & 16454\\
HST & 2020-12-12 & 13:43:09 & Wong & 16454\\
HST & 2020-12-12 & 13:44:51 & Wong & 16454\\
HST & 2020-12-12 & 13:46:33 & Wong & 16454\\
HST & 2020-12-12 & 13:48:15 & Wong & 16454\\
HST & 2020-12-12 & 16:15:43 & Wong & 16454\\
HST & 2020-12-12 & 16:17:25 & Wong & 16454\\
HST & 2020-12-12 & 16:19:07 & Wong & 16454\\
HST & 2020-12-12 & 16:20:49 & Wong & 16454\\
HST & 2020-12-12 & 16:22:31 & Wong & 16454\\
HST & 2020-12-12 & 16:24:13 & Wong & 16454\\
HST & 2020-12-12 & 16:27:41 & Wong & 16454\\
HST & 2020-12-12 & 16:31:09 & Wong & 16454\\
HST & 2020-12-12 & 16:42:41 & Wong & 16454\\
HST & 2020-12-12 & 16:48:15 & Wong & 16454\\
HST & 2020-12-12 & 16:53:49 & Wong & 16454\\
HST & 2020-12-12 & 16:55:31 & Wong & 16454\\
HST & 2020-12-12 & 16:57:13 & Wong & 16454\\
HST & 2020-12-12 & 16:58:55 & Wong & 16454\\
HST & 2020-12-12 & 19:26:22 & Wong & 16454\\
HST & 2020-12-12 & 19:28:04 & Wong & 16454\\
HST & 2020-12-12 & 19:29:46 & Wong & 16454\\
HST & 2020-12-12 & 19:31:28 & Wong & 16454\\
HST & 2020-12-12 & 19:33:10 & Wong & 16454\\
HST & 2020-12-12 & 19:34:52 & Wong & 16454\\
HST & 2020-12-12 & 19:38:20 & Wong & 16454\\
HST & 2020-12-12 & 19:41:48 & Wong & 16454\\
HST & 2020-12-12 & 19:53:20 & Wong & 16454\\
HST & 2020-12-12 & 19:58:54 & Wong & 16454\\
HST & 2020-12-12 & 20:04:28 & Wong & 16454\\
HST & 2020-12-12 & 20:06:10 & Wong & 16454\\
HST & 2020-12-12 & 20:07:52 & Wong & 16454\\
HST & 2020-12-12 & 20:09:34 & Wong & 16454\\
\enddata
\end{deluxetable}

\bibliography{references}{}
\bibliographystyle{aasjournal}

\end{document}